%% file: Paper-dim6.tex
\documentclass[a4paper, 10pt]{article}

\usepackage{jheppub}

\usepackage{amsthm}

\usepackage[numbers,sort&compress]{natbib}
\usepackage{dsfont}
\usepackage{slashed}
\usepackage{color}
\usepackage{empheq}

\usepackage{adjustbox}

\usepackage{bibgerm}

\usepackage{soul}

\usepackage[english]{babel}

\usepackage[utf8x]{inputenc}

\usepackage{graphicx,epsfig}
\usepackage{pstricks}

\usepackage{tikz}
\usepackage{tikz-feynman}

\usepackage{bashful}

\usepackage{subfigure}

\usepackage{setspace}

\usepackage{booktabs}

\usepackage{float}

\usepackage{nextpage}

\usepackage{pdfpages}

\usepackage{hypcap}

\usepackage[small,bf]{caption}

\usepackage{longtable}

\usepackage{fancyhdr}

\usepackage[makeroom]{cancel}

\usepackage{microtype}

\usepackage{feynmp-auto-mod}

\usepackage{xparse}
\usepackage{etoolbox}

\usepackage[subfigure]{tocloft}

\usepackage{listings}

\newcommand{\p}{\partial}

\renewcommand{\Im}{\mathrm{Im}}

\newcommand{\<}{\langle}
\renewcommand{\>}{\rangle}

\renewcommand{\O}{\mathcal{O}}

\newcommand{\tr}{\mathrm{Tr}}

\renewcommand{\L}{\mathcal{L}}

\newcommand{\E}{\mathcal{E}}

\newcommand{\nn}{\nonumber\\}

\newcommand{\q}{\mathsf{q}}

\newcommand{\msbar}{$\overline{\text{MS}}$}

\NewDocumentCommand{\g}{ O{} }{
	\ifblank{#1}{\bar g}{\bar g_{#1}}
}

\NewDocumentCommand{\Op}{ m m O{} o }{
	\O^{\ifblank{#3}{}{#3,}#2 }_{\IfNoValueTF{#4}{#1}{\substack{#1\\#4}}}
}
\NewDocumentCommand{\EOp}{ m m O{} o }{
	\E^{\ifblank{#3}{}{#3,}#2 }_{\IfNoValueTF{#4}{#1}{\substack{#1\\#4}}}
}
\NewDocumentCommand{\lwc}{ m m O{} o }{
	L^{\ifblank{#3}{}{#3,}#2 }_{\IfNoValueTF{#4}{#1}{\substack{#1\\#4}}}
}
\NewDocumentCommand{\kwc}{ m m O{} o }{
	K^{\ifblank{#3}{}{#3,}#2 }_{\IfNoValueTF{#4}{#1}{\substack{#1\\#4}}}
}
\NewDocumentCommand{\dlwc}{ m m O{} o }{
	{\dot L}^{\ifblank{#3}{}{#3,}#2 }_{\IfNoValueTF{#4}{#1}{\substack{#1\\#4}}}
}
\NewDocumentCommand{\cwc}{ m m O{} o }{
	C^{\ifblank{#3}{}{#3,}#2 }_{\IfNoValueTF{#4}{#1}{\substack{#1\\#4}}}
}

\newcommand{\XXint}[3]{{\setbox0=\hbox{$#1{#2#3}{\int}$}
\vcenter{\hbox{$#2#3$ }}\kern-.65\wd0}}

\newcommand{\remark}[1]{}

\definecolor{darkgreen}{rgb}{0,0.5,0}
\definecolor{darkblue}{rgb}{0,0,0.5}
\definecolor{darkred}{rgb}{0.5,0,0}
\definecolor{beige}{rgb}{0.7,0.4,0.3}

\makeatletter
  \def\my@tag@font{\normalsize}
  \def\maketag@@@#1{\hbox{\m@th\normalfont\my@tag@font#1}}
  \let\amsmath@eqref\eqref
  \renewcommand\eqref[1]{{\let\my@tag@font\relax\amsmath@eqref{#1}}}
\makeatother

\newcommand{\roverline}[1]{\mathpalette\doroverline{#1}}
\newcommand{\doroverline}[2]{\overline{#1#2}}

\newenvironment{myfmf}[1]
{\begin{fmffile}{#1}
\fmfcmd{%
  style_def wboson expr p =
  cdraw (wiggly p);
  shrink (1);
  cfill (arrow p);
  endshrink;
  enddef;}
\fmfcmd{%
  style_def momins expr p =
  drawarrow p;
  enddef;}
\fmfcmd{
    path quadrant, q[], otimes;
    quadrant = (0, 0) -- (0.5, 0) & quartercircle & (0, 0.5) -- (0, 0);
    for i=1 upto 4: q[i] = quadrant rotated (45 + 90*i); endfor
    otimes = q[1] & q[2] & q[3] & q[4] -- cycle;
}
\fmfwizard
  }
{
\end{fmffile} 
}

\makeatletter
\renewcommand\paragraph{\@startsection{paragraph}{4}{\z@}%
  {-3.25ex\@plus -1ex \@minus -.2ex}%
  {1.5ex \@plus .2ex}%
  {\normalfont\normalsize\bfseries}}
\makeatother

\cftsetindents{section}{0em}{2em}
\cftsetindents{subsection}{2em}{3em}
\cftsetindents{subsubsection}{5em}{4em}

\allowdisplaybreaks[1]

\preprint{
\mbox{}\hfill{} ZU-TH 48/25
}

\title{\boldmath Renormalization-group equations of the LEFT at two loops: dimension-six operators}

\author{Luca Naterop,}

\author{Peter Stoffer}

\emailAdd{luca.naterop@physik.uzh.ch}
\emailAdd{stoffer@physik.uzh.ch}

\affiliation{Physik-Institut, Universit\"at Z\"urich, Winterthurerstrasse 190, 8057 Z\"urich, Switzerland}
\affiliation{PSI Center for Neutron and Muon Sciences, 5232 Villigen PSI, Switzerland}

\abstract{
	We present the third part of a systematic calculation of the two-loop anomalous dimensions for the low-energy effective field theory below the electroweak scale (LEFT): insertions of dimension-six operators that conserve baryon number. In line with our previous publications, we obtain the results in the algebraically consistent 't Hooft--Veltman scheme for $\gamma_5$, corrected for evanescent as well as chiral-symmetry-breaking effects through finite renormalizations. We compute the renormalization of the dimension-six four-fermion and three-gluon operators, as well as the power corrections to lower-dimension operators in the presence of masses, i.e., the down-mixing into dimension-five dipole operators, masses, gauge couplings, and theta terms. Our results are of interest for a broad range of low-energy precision searches for physics beyond the Standard Model.
}

\numberwithin{equation}{section}

\begin{document}

	\maketitle

		
	\begin{myfmf}{diags/diags}
	
	\input{sections/Introduction}
	\input{sections/LEFT}

	\input{sections/Computation}
	\input{sections/Results}

	\input{sections/Conclusions}

	\section*{Note added}
	
	Shortly after the first version of the present work Ref.~\cite{DiNoi:2025arz} appeared, which considers the two-loop QCD mixing of vector-type four-quark operators into the top mass in SMEFT and has some partial overlap with the present work.
	
	\section*{Acknowledgements}
	\addcontentsline{toc}{section}{\numberline{}Acknowledgements}

	We thank J.~de Vries, G.~Falcioni, J.~Fuentes-Mart\'in, B.~Grinstein, R.~Gr\"ober, F.~Herzog, \linebreak A.~V.~Manohar. M.~Misiak, S.~Di~Noi, B.~Ruijl, C.-H.~Shen, A.~Signer, D.~St\"ockinger, A.~E.~Thomsen, and M.~Zoller for useful discussions.
	Financial support by the Swiss National Science Foundation (Project No.~PCEFP2\_194272) is gratefully acknowledged.

	
	\appendix

	\input{sections/ConventionsSupplement}

	\input{sections/BasisChange}

	\end{myfmf}

	\addcontentsline{toc}{section}{\numberline{}References}
	\bibliographystyle{utphysmod}
	\bibliography{Literature}
	
\end{document}

%% file: sections/Introduction.tex

\section{Introduction}

The application of perturbation theory to problems with multiple scales leads to logarithms of the ratio of scales. In the case of widely separated scales, these logarithms become large and can spoil the perturbative expansion, as they multiply the couplings that are used as expansion parameters of a fixed-order calculation. This problem can be solved by rearranging the expansion and resumming large logarithms, which leads to a leading-log expansion instead of a fixed-order expansion. The resummation is achieved by solving the renormalization-group equations (RGEs) of the theory. In order to resum all logarithms, the heavy scales need to be removed from the theory or integrated out, which leads to an effective field theory (EFT) valid at energy scales below the heavy scale that was integrated out. While a fixed-order calculation in the full theory suffers from the large scale separation, the EFT takes advantage of it: it provides a systematic expansion in the ratio of scales, which is a small dimensionless parameter. The concept of EFTs has a wide range of applications in particle physics and beyond. In particular, it has gained significant attention in past years in the context of searches for physics beyond the Standard Model (SM). The absence of signals of new physics in collider searches suggests that new particles beyond the SM could be very heavy, which leads to a separation between the scale of new physics and the electroweak scale. Under the assumption of linear realization of electroweak symmetry, this scale separation enables a description of indirect effects beyond the SM in terms of the SMEFT~\cite{Buchmuller:1985jz,Grzadkowski:2010es}, which remains largely agnostic about the details of heavy new physics, but can be matched to a variety of explicit models in the ultraviolet (UV).

For precision observables at very low energies, the low-energy effective field theory below the electroweak scale (LEFT) should be used, which takes advantage of the scale separation between the heavy SM particles (top quark, Higgs boson, and the electroweak gauge bosons $W^\pm$ and $Z$) and all the other degrees of freedom in the SM. The LEFT is the generalization of Fermi's theory of weak interaction and it is the most general theory below the electroweak scale respecting QCD and QED gauge symmetries. It describes the low-energy effects not only of electroweak physics but also of any heavy degrees of freedom beyond the SM.

Increased experimental precision in low-energy searches calls for a theory framework at an accuracy that does not dilute the experimental sensitivities when connecting the low-energy scale to the heavy scale of new physics. The EFT framework consisting of SMEFT and LEFT is currently fully worked out at leading-log accuracy: the operator bases are known to high dimensions~\cite{Buchmuller:1985jz,Grzadkowski:2010es,Lehman:2014jma,Liao:2016hru,Murphy:2020rsh,Li:2020gnx,Liao:2020jmn,Harlander:2023psl,Jenkins:2017jig,Liao:2020zyx,Murphy:2020cly,Li:2020tsi} and the complete one-loop renormalization of the two theories up to dimension six was performed in Refs.~\cite{Jenkins:2013zja,Jenkins:2013wua,Alonso:2013hga,Jenkins:2017dyc}. While partial results are known to high loop orders~\cite{Buras:1989xd,Buras:1991jm,Buras:1992tc,Ciuchini:1993vr,Ciuchini:1993fk,Buchalla:1995vs,Chetyrkin:1997gb,Buras:2000if,Bobeth:2003at,Gorbahn:2004my,Huber:2005ig,Gorbahn:2005sa,Czakon:2006ss,Aebischer:2017gaw,Panico:2018hal,Morell:2024aml}, a broad effort is ongoing to advance this EFT framework to complete next-to-leading-log accuracy. In addition to one-loop matrix elements, this entails the calculation of the one-loop matching of UV models to SMEFT, which has been largely automated~\cite{Carmona:2021xtq,Fuentes-Martin:2022jrf,Fuentes-Martin:2023ljp,Aebischer:2023nnv,Thomsen:2024abg}, the one-loop matching between SMEFT and LEFT~\cite{Dekens:2019ept}, as well as the two-loop RGEs of SMEFT and LEFT~\cite{Gorbahn:2016uoy,deVries:2019nsu,Bern:2020ikv,Aebischer:2022anv,Fuentes-Martin:2022vvu,Aebischer:2023djt,Jenkins:2023rtg,Jenkins:2023bls,Naterop:2023dek,DiNoi:2023ygk,Fuentes-Martin:2023ljp,Aebischer:2024xnf,Manohar:2024xbh,DiNoi:2024ajj,Born:2024mgz,Naterop:2024cfx,Fuentes-Martin:2024agf,Aebischer:2025hsx,Duhr:2025zqw,Haisch:2025lvd,Naterop:2025lzc,Haisch:2025vqj,Duhr:2025yor,Banik:2025wpi}.

An example of low-energy precision observables calling for higher-order calculations are searches for $CP$-violating electric dipole moments, see, e.g., Ref.~\cite{Alarcon:2022ero} for a review. Calculations in the $CP$-odd sector of the EFTs involve $\gamma_5$ or the Levi-Civita symbol, which are objects strictly defined in four space-time dimensions, leading to well-known obstacles with dimensional regularization~\cite{Jegerlehner:2000dz}. The only scheme known to be algebraically consistent to any loop order is the original scheme introduced by 't~Hooft and Veltman (HV)~\cite{tHooft:1972tcz,Breitenlohner:1975hg,Breitenlohner:1976te,Breitenlohner:1977hr}, also known as Breitenlohner--Maison/'t~Hooft--Veltman (BMHV) scheme. Compared to naive dimensional regularization (NDR), its consistency comes at the expense of higher computational complexity, more complicated evanescent structures, as well as the spurious breaking of chiral symmetries, which can be restored by finite counterterms~\cite{Schubert:1988ke,Ferrari:1994ct,Cornella:2022hkc,Belusca-Maito:2020ala,Belusca-Maito:2021lnk,Belusca-Maito:2023wah,Stockinger:2023ndm,OlgosoRuiz:2024dzq,Ebert:2024xpy,DiNoi:2025uan,vonManteuffel:2025swv,Fuentes-Martin:2025meq}.

In Ref.~\cite{Naterop:2023dek}, we established a HV scheme for the LEFT at dimension six that incorporates two types of finite renormalization. First, the one-loop renormalization of evanescent operators ensures a separation of the physical and evanescent sectors. Second, finite counterterms restore global chiral symmetry of the LEFT in the spurion sense. Based on this scheme definition, we are computing the complete two-loop RGEs of the LEFT: in Ref.~\cite{Naterop:2024cfx}, we considered all dimension-five effects, and in Ref.~\cite{Naterop:2025lzc}, we treated the baryon-number-violating sector at dimension six. In the present article, we extend this work to the baryon-number-conserving operators at dimension six. We compute the complete two-loop RGEs for the mixing among four-fermion and three-gluon operators, as well as the power-suppressed down-mixing into lower-dimension operators in the presence of masses. In the baryon-number-violating sector, the two-loop RGEs can also be computed in the NDR scheme without encountering issues with $\gamma_5$~\cite{Aebischer:2025hsx,Naterop:2025lzc}. The two-loop running and mixing among all four-fermion operators in the LEFT was recently reconstructed in the NDR scheme~\cite{Aebischer:2025hsx} from known results for UV poles: the $\gamma_5$ problem could be circumvented by making use of one-loop basis changes. This method does not work for all down-mixings into lower-dimension operators, as in this case ill-defined $\gamma_5$-odd traces show up already in finite one-loop contributions. An alternative to the HV scheme could be to give up the cyclicity of the trace~\cite{Kreimer:1989ke,Korner:1991sx,Chen:2023lus,Chen:2024hlv}, but to the best of our knowledge there is no proof of the consistency of such a prescription that is valid for EFTs.

The rest of the article is structured as follows. In Sect.~\ref{sec:LEFT}, we recall our conventions for the LEFT and we discuss off-shell renormalization and field redefinitions. In Sect.~\ref{sec:Computation}, we explain some details of our calculations, whereas in Sect.~\ref{sec:Results} we discuss selected aspects of our results, such as scheme dependences and the results for the insertions of three-gluon operators. The complete set of two-loop RGEs due to the insertion of dimension-six operators is provided as supplementary material.

%% file: sections/LEFT.tex

\section{LEFT}
\label{sec:LEFT}

\subsection{Off-shell renormalization}

The LEFT is defined as the EFT consisting of QCD and QED, supplemented by a tower of higher-dimension operators that respect $SU(3)_c \times U(1)_\mathrm{em}$ gauge invariance and are suppressed by powers of the vacuum expectation value $v$. The Lagrangian is
\begin{equation}
	\label{eq:LEFTLagrangian}
	\L_\mathrm{LEFT} = \L_\mathrm{QCD+QED} + \L_{\nu} + \sum_i L_i \O_i + \sum_i L_i^\text{red} \O_i^\text{red} + \sum_i K_i \E_i \, ,
\end{equation}
with the dimension-four QCD and QED Lagrangian  
\begin{align}
	\label{eq:qcdqed}
	\L_{\rm QCD + QED} &= - \frac14 G_{\mu \nu}^A G^{A \mu \nu} -\frac14 F_{\mu \nu} F^{\mu\nu} + \theta_{\rm QCD} \frac{g^2}{32 \pi^2} G_{\mu \nu}^A \widetilde G^{A \mu \nu} +  \theta_{\rm QED} \frac{e^2}{32 \pi^2} F_{\mu \nu} \widetilde F^{\mu \nu} \nn
		&\quad + \sum_{\psi=u,d,e}\overline \psi \left( i \slashed D - M_\psi P_L - M_\psi^\dagger P_R \right) \psi + \L_\mathrm{gf}^\mathrm{QCD+QED} + \L_\mathrm{FP}^\mathrm{QCD} \, ,
\end{align}
the neutrino Lagrangian including a lepton-number-violating Majorana mass term
\begin{equation}
	\L_\nu = \bar\nu_L i \slashed \p \nu_L - \frac{1}{2} \left( \nu_L^T C M_\nu \nu_L + \bar\nu_L M_\nu^\dagger C \bar\nu_L^T \right) \, ,
\end{equation}
and physical operators $\O_i$ with coefficients $L_i$. The covariant derivative is defined as $D_\mu = \p_\mu + i g T^A G_\mu^A + i e Q A_\mu$. We perform the renormalization off shell, which induces the presence of redundant operators $\O_i^\text{red}$ that are related to the physical operators via the classical equations of motion (EOM) and can be removed through field redefinitions. In dimensional regularization, the renormalization also generates evanescent operators $\E_i$ with coefficients $K_i$, which vanish in four dimensions but need to be taken into account when calculating finite one-loop terms as well as two-loop divergences. The exact definition of the different types of physical and evanescent operators is part of the scheme: the complete operator basis to dimension six, including on-shell redundant as well as evanescent operators in the HV scheme is provided in Ref.~\cite{Naterop:2023dek}. In particular, we define all physical higher-dimension operators to contain only index summations over four space-time dimensions, whereas summations over indices in $D-4 = -2\varepsilon$ dimensions are part of the evanescent sector.

Due to the need of gauge fixing in perturbative calculations, gauge symmetry is broken to BRST symmetry, which in general gives rise to additional unphysical gauge-variant (but BRST-exact) nuisance operators. Considering these gauge-variant operators explicitly can be avoided by making use of the background-field method~\cite{Abbott:1980hw,Abbott:1983zw}. Our conventions for the gauge-fixing and ghost Lagrangians $\L_\mathrm{GF}$ and $\L_\mathrm{FP}$ are given in Ref.~\cite{Naterop:2023dek}. The gauge of the background fields does not need be fixed in an off-shell calculation of the one-particle-irreducible (1PI) Green's functions. Therefore, manifest gauge invariance with respect to the background fields is preserved. Although sub-divergences of two-loop diagrams involve quantum fields and hence give rise to gauge-variant counterterms, the explicit construction of the gauge-variant nuisance operators can be avoided in the calculation of the two-loop counterterms to physical operators, either by making use of the $\bar R$-operation or by employing a variant of the infrared rearrangement, see Ref.~\cite{Naterop:2024cfx} for details. Here, we make use of the $\bar R$-operation in combination with an auxiliary mass as IR regulator~\cite{Chetyrkin:1997fm}. Although the auxiliary mass introduces even BRST-violating sub-divergences, the $\bar R$-operation automatically subtracts them, leaving the overall UV divergence unchanged. We compute all divergent two-loop counterterms to the coefficients of physical and redundant operators, $L_i$ and $L_i^\mathrm{red}$, by computing the necessary 1PI Green's functions with insertions of physical dimension-six operators. We do not consider insertions of evanescent operators in two-loop diagrams and we disregard the two-loop counterterms to the coefficients of evanescent operators: both effects become relevant only beyond next-to-leading-log accuracy.

We find that the divergent two-loop counterterms to physical operators are independent of the QED gauge parameter $\xi_\gamma$. However, in the case of insertions of the three-gluon operator $\O_G$, we find that the off-shell counterterms depend on the QCD gauge parameter $\xi_g$. As expected, this gauge-parameter dependence drops out once we remove the redundant operators through field redefinitions.

\subsection{Field redefinitions}

The physical parameters of the theory are obtained after removing the redundant operators $\O_i^\text{red}$ through field redefinitions. We make use of non-linear field redefinitions as discussed in detail in Ref.~\cite{Naterop:2023dek} for the one-loop case. In particular, up to dimension-six effects the fermion fields can be redefined as~\cite{Dekens:2019ept,Naterop:2023dek}\footnote{The bar denotes the restriction of Lorentz-index summations over four space-time dimensions. The effects of the two-loop field redefinitions on the evanescent sector are not relevant for the present work.}
\begin{align}
	\label{eq:FermionFieldRedefinition}
	\psi_{L,R} &\mapsto \psi_{L,R} + A_{L,R}^\psi \psi_{L,R} + B_{L,R}^\psi i \bar{\slashed D} \psi_{R,L} + C_{L,R}^\psi (i \bar{\slashed D})^2 \psi_{L,R} \nn
		&\quad + D_{L,R}^{\psi\gamma} \bar\sigma^{\mu\nu} F_{\mu\nu} \psi_{L,R} + D_{L,R}^{\psi g} \bar\sigma^{\mu\nu} G_{\mu\nu} \psi_{L,R} \, ,
\end{align}
where $A_{L,R}$, $B_{L,R}$, $C_{L,R}$, and $D_{L,R}^{\gamma,g}$ are matrices in flavor space and the term involving $G_{\mu\nu}$ is only present for quarks. The neutrino field redefinition reads
\begin{align}
	\label{eq:NeutrinoFieldRedefinition}
	\nu_L &\mapsto \nu_L + A^\nu \nu_L + B^\nu i \bar{\slashed \p} \nu_R + C^\nu (i \bar{\slashed \p})^2 \nu_L + D^{\nu\gamma} \bar\sigma^{\mu\nu} F_{\mu\nu} \nu_L \, , \nn
	\nu_R &\mapsto \nu_R + (A^\nu)^* \nu_R + (B^\nu)^* i \bar{\slashed \p} \nu_L + (C^\nu)^* (i \bar{\slashed \p})^2 \nu_R - (D^{\nu\gamma})^* \bar\sigma^{\mu\nu} F_{\mu\nu} \nu_R \, .
\end{align}
The redefinition of gauge fields up to dimension-six effects is given by
\begin{align}
	\label{eq:GaugeFieldRedefinition}
	A_\mu &\mapsto A_\mu + b^\gamma \, \p^\nu F_{\nu\mu} + \sum_{\psi=u,d,e,\nu} \bar\psi_L C_L^{\gamma\psi} \bar\gamma_\mu \psi_L + \sum_{\psi=u,d,e} \bar\psi_R C_R^{\gamma\psi} \bar\gamma_\mu \psi_R \, , \nn
	G_\mu^A &\mapsto G_\mu^A + b^g \, (D^\nu G_{\nu\mu})^A + \sum_{\psi=u,d} \left( \bar\psi_L C_L^{g\psi} \bar\gamma_\mu T^A \psi_L + \bar\psi_R C_R^{g\psi} \bar\gamma_\mu T^A \psi_R \right) \, ,
\end{align}
where $C_{L,R}^{\gamma\psi}$, $C_{L,R}^{g\psi}$ are again matrices in flavor space. These field redefinition are restricted only by the LEFT power counting and have the same generic form at any loop order. At two loops, one needs to take into account the product of two one-loop effects. As an example, consider the field redefinition for the photon field. The coefficient $b^\gamma$ and the matrices $C_{L,R}^{\gamma\psi}$ are chosen such that the redundant operators related to classical EOM for the photon field
\begin{equation}
	\p_\mu F^{\mu\nu} = e j_\mathrm{em}^\nu \, , \quad j_\mathrm{em}^\mu = \sum_{\psi=u,d,e} \q_\psi \, \bar\psi \gamma^\mu \psi
\end{equation}
are removed from the Lagrangian. This leads to
\begin{equation}
	b_\gamma = - \lwc{\gamma D}{} \, , \quad c_L^{\gamma\psi} = - e  \q_\psi \lwc{\gamma D}{} - \lwc{\psi\gamma D}{L} \, , \quad c_R^{\gamma\psi} = - e \q_\psi \lwc{\gamma D}{} - \lwc{\psi\gamma D}{R} 
\end{equation}
in terms of the coefficients of the redundant operators
\begin{equation}
	\Op{\gamma D}{} = \overline{\p_\mu F^{\mu\nu} \p^\lambda F_{\lambda\nu}} \, , \quad \Op{\psi\gamma D}{L} = (\bar\psi_L \bar\gamma_\nu \psi_L) (\bar\p_\mu F^{\mu\nu}) \, , \quad \Op{\psi\gamma D}{R} = (\bar\psi_R \bar\gamma_\nu \psi_R) (\bar\p_\mu F^{\mu\nu}) \,.
\end{equation}
These field redefinitions should be understood as acting on the bare Lagrangian, i.e., they remove the entire redundant operators. Splitting the bare parameters into renormalized parameters and counterterms, this implies for the divergent two-loop shifts, e.g.,
\begin{equation}
	\label{eq:2LoopFieldRed}
	c_L^{\gamma\psi}\Big|_\text{2-loop} = - \delta_\mathrm{1L}(e) \q_\psi \delta_{1L}(\lwc{\gamma D}{}) - e \q_\psi \delta_{2L}(\lwc{\gamma D}{}) - \delta_{2L}(\lwc{\psi\gamma D}{L}) \, ,
\end{equation}
where $\delta_{1L}$ and $\delta_{2L}$ denote one- and two-loop counterterms and the renormalized coefficients of redundant operators are assumed to vanish. The first term in Eq.~\eqref{eq:2LoopFieldRed} consists of a product of two one-loop counterterms. Its omission would lead to an incorrect result for the counterterms after field redefinition, in particular it would induce an apparent violation of the 't~Hooft consistency conditions~\cite{tHooft:1972tcz,tHooft:1973mfk}, which is a useful check of the $1/\varepsilon^2$ poles of the two-loop counterterms. In the fermion sector, similar effects also appear at $\O(1/\varepsilon)$.

We denote the counterterms to a parameter $X_i$ by
\begin{equation}
	\label{eq:CountertermEpsilonExpansion}
	X_i^\mathrm{ct} = \sum_{l=1}^\infty \sum_{n=0}^l \frac{1}{\varepsilon^n} \frac{1}{(16\pi^2)^l} X_i^{(l,n)}(L_j^r,K_j^r) \, ,
\end{equation}
where $l$ is the loop order and $L_j^r$, $K_j^r$ are the renormalized physical and evanescent parameters.\footnote{For simplicity, we will often drop the superscript ${}^r$.} Consider the contribution of an insertion of $\Op{\nu d}{LL}[V][]$ to the counterterm for $\lwc{\nu e}{LL}[V][]$. The two-loop contribution to the $1/\varepsilon^2$ counterterm that scales as $\O(N_c^2 n_u \q_u^2 \q_d \q_e)$ is given after field redefinition by
\begin{equation}
	\label{eq:4FOnShellExample}
	\Big(\lwc{\nu e}{LL}[V][prst]\Big)^{(2,2)} = \frac{8}{9} e^4 N_c^2 n_u \q_u^2 \q_d \q_e \delta_{st} \lwc{\nu d}{LL}[V][prww] + \ldots \, ,
\end{equation}
where $p$, $r$, $s$, $t$, $w$ denote flavor indices and a sum over $w$ is implicit. This counterterm can be obtained in an on-shell calculation from the connected four-point diagrams including counterterms
\begin{equation}
	\begin{gathered}
		\begin{fmfgraph*}(60,100)
			\fmfset{curly_len}{2mm}
			\fmftop{t1,t2} \fmfbottom{b1,b2}
			\fmf{quark,tension=5}{t1,v1}
			\fmf{quark,tension=5,label.side=left,label=$e$}{v1,t2}
			\fmf{photon,tension=3}{v1,v2}
			\fmf{quark,right}{v2,v3}
			\fmf{quark,right,label=$u$}{v3,v2}
			\fmf{photon,tension=3}{v3,v4}
			\fmf{quark,right}{v4,v5}
			\fmf{quark,right,label=$d$}{v5,v4}
			\fmf{phantom,tension=7}{v5,v6}
			\fmf{quark,tension=5}{b1,v6}
			\fmf{quark,tension=5,label=$\nu$,label.side=right}{v6,b2}
			\fmfdot{v5,v6}
		\end{fmfgraph*}
	\end{gathered}
	\quad + \quad
	\begin{gathered}
		\begin{fmfgraph*}(60,100)
			\fmfset{curly_len}{2mm}
			\fmftop{t1,t2} \fmfbottom{b1,b2}
			\fmf{quark,tension=5}{t1,v1}
			\fmf{quark,tension=5,label.side=left,label=$e$}{v1,t2}
			\fmf{photon,tension=2}{v1,v2}
			\fmf{quark,right}{v2,v3}
			\fmf{quark,right,label=$u$}{v3,v2}
			\fmf{phantom,tension=7}{v3,v6}
			\fmf{quark,tension=5}{b1,v6}
			\fmf{quark,tension=5,label=$\nu$,label.side=right}{v6,b2}
			\fmfv{decor.shape=otimes, decor.filled=empty, decor.size=(2.5mm)}{v3,v6}
		\end{fmfgraph*}
	\end{gathered}
	\quad + \quad
	\begin{gathered}
		\begin{fmfgraph*}(60,100)
			\fmfset{curly_len}{2mm}
			\fmftop{t1,t2} \fmfbottom{b1,b2}
			\fmf{quark,tension=5}{t1,v1}
			\fmf{quark,tension=5,label.side=left,label=$e$}{v1,t2}
			\fmf{photon,tension=2}{v1,v2}
			\fmf{photon,tension=2}{v2,v4}
			\fmf{quark,right}{v4,v5}
			\fmf{quark,right,label=$d$}{v5,v4}
			\fmf{phantom,tension=8}{v5,v6}
			\fmf{quark,tension=5}{b1,v6}
			\fmf{quark,tension=5,label=$\nu$,label.side=right}{v6,b2}
			\fmfdot{v5,v6}
			\fmfv{decor.shape=otimes, decor.filled=empty, decor.size=(3mm)}{v2}
		\end{fmfgraph*}
	\end{gathered}
	\quad + \quad
	\begin{gathered}
		\begin{fmfgraph*}(60,100)
			\fmfset{curly_len}{2mm}
			\fmftop{t1,t2} \fmfbottom{b1,b2}
			\fmf{quark,tension=1}{t1,v1}
			\fmf{quark,tension=1,label.side=left,label=$e$}{v1,t2}
			\fmf{phantom,tension=8}{v1,v2}
			\fmf{quark,tension=1}{b1,v2}
			\fmf{quark,tension=1,label=$\nu$,label.side=right}{v2,b2}
			\fmfv{decor.shape=square, decor.filled=hatched, decor.size=(2mm)}{v1,v2}
		\end{fmfgraph*}
	\end{gathered}
	\quad = \text{finite} \, ,
\end{equation}
where crossed circles stand for one-loop and hatched boxes for two-loop counterterms. In an off-shell calculation, there is no corresponding two-loop diagram, as only 1PI diagrams are considered. At one loop, the counterterms are determined off shell by
\begin{equation}
	\begin{gathered}
		\begin{fmfgraph*}(70,40)
			\fmfset{curly_len}{2mm}
			\fmfleft{l1} \fmfright{r1}
			\fmf{photon,tension=2}{l1,v1}
			\fmf{quark,right}{v1,v2}
			\fmf{quark,right,label=$u$,label.side=right}{v2,v1}
			\fmf{photon,tension=2}{v2,r1}
		\end{fmfgraph*}
	\end{gathered}
	\quad + \quad
	\begin{gathered}
		\begin{fmfgraph*}(70,40)
			\fmfset{curly_len}{2mm}
			\fmfleft{l1} \fmfright{r1}
			\fmf{photon,label=$$,tension=2}{l1,v1,r1}
			\fmfv{decor.shape=otimes, decor.filled=empty, decor.size=(3mm)}{v1}
		\end{fmfgraph*}
	\end{gathered}
	\quad = \text{finite}
\end{equation}
and
\begin{equation}
	\begin{gathered}
		\begin{fmfgraph*}(70,70)
			\fmfset{curly_len}{2mm}
			\fmftop{t1} \fmfbottom{b1,b2}
			\fmf{photon,tension=2}{t1,v2}
			\fmf{quark,right}{v2,v3}
			\fmf{quark,right,label=$d$}{v3,v2}
			\fmf{phantom,tension=8}{v3,v6}
			\fmf{quark,tension=5}{b1,v6}
			\fmf{quark,tension=5,label=$\nu$,label.side=right}{v6,b2}
			\fmfdot{v3,v6}
		\end{fmfgraph*}
	\end{gathered}
	\quad + \quad
	\begin{gathered}
		\begin{fmfgraph*}(70,60)
			\fmfset{curly_len}{2mm}
			\fmftop{t1} \fmfbottom{b1,b2}
			\fmf{photon,tension=3}{t1,v2}
			\fmf{quark,tension=5}{b1,v2}
			\fmf{quark,tension=5,label=$\nu$,label.side=right}{v2,b2}
			\fmfv{decor.shape=otimes, decor.filled=empty, decor.size=(3mm)}{v2}
		\end{fmfgraph*}
	\end{gathered}
	\quad = \text{finite} \, ,
\end{equation}
giving
\begin{align}
	\delta Z_\gamma^{1/2} &= - \frac{e^2 N_c n_u \q_u^2}{24\pi^2 \varepsilon} + \ldots \, , \quad e^{(1,1)} = \frac{2}{3} e^3 N_c n_u \q_u^2 + \ldots \, , \nn
	\Big( \lwc{\nu\gamma D}{L}[][pr] \Big)^{(1,1)} &= \frac{2}{3} e N_c \q_d \lwc{\nu d}{LL}[V][prww] + \ldots \, .
\end{align}
At two loops, the only 1PI contributions are counterterms and wave-function renormalization
\begin{equation}
	\begin{gathered}
		\begin{fmfgraph*}(70,60)
			\fmfset{curly_len}{2mm}
			\fmftop{t1} \fmfbottom{b1,b2}
			\fmf{photon,tension=3}{t1,v2}
			\fmf{quark,tension=5}{b1,v2}
			\fmf{quark,tension=5,label=$\nu$,label.side=right}{v2,b2}
			\fmfv{decor.shape=otimes, decor.filled=empty, decor.size=(3mm)}{v2}
		\end{fmfgraph*}
	\end{gathered} \!\!\!\! \times \delta Z_\gamma^{1/2}
	\quad + \quad
	\begin{gathered}
		\begin{fmfgraph*}(70,60)
			\fmfset{curly_len}{2mm}
			\fmftop{t1} \fmfbottom{b1,b2}
			\fmf{photon,tension=3}{t1,v2}
			\fmf{quark,tension=5}{b1,v2}
			\fmf{quark,tension=5,label=$\nu$,label.side=right}{v2,b2}
			\fmfv{decor.shape=square, decor.filled=hatched, decor.size=(3mm)}{v2}
		\end{fmfgraph*}
	\end{gathered}
	\quad = \text{finite} \, ,
\end{equation}
leading to the off-shell counterterm
\begin{equation}
	\Big(\lwc{\nu\gamma D}{L}[][pr]\Big)^{(2,2)} = \frac{4}{9} e^3 N_c^2 n_u \q_u^2 \q_d \lwc{\nu d}{LL}[V][prww] + \ldots \, .
\end{equation}
Performing the field redefinition~\eqref{eq:GaugeFieldRedefinition} leads to a shift of the four-fermion coefficient
\begin{equation}
	\Big( \lwc{\nu e}{LL}[V][prst] \Big)^{(2,2)} \mapsto \Big( \lwc{\nu e}{LL}[V][prst] \Big)^{(2,2)} + e^{(1,1)} \q_e \delta_{st} \, \Big( \lwc{\nu\gamma D}{L}[][pr] \Big)^{(1,1)} + e \q_e \delta_{st} \, \Big( \lwc{\nu\gamma D}{L}[][pr] \Big)^{(2,2)} \, ,
\end{equation}
which reproduces Eq.~\eqref{eq:4FOnShellExample}, provided that the product of one-loop terms is correctly taken into account. Similar considerations apply to the redefinitions of gluon and fermion fields.

\subsection{RGEs and consistency condition}

In terms of the counterterms~\eqref{eq:CountertermEpsilonExpansion}, the RGEs in the presence of finite renormalizations are given by~\cite{Naterop:2023dek}
\begin{align}
	\label{eq:RGEMasterFormula}
	\frac{d L_i^r(\mu)}{d\log\mu} &= \frac{1}{16\pi^2} 2 L_i^{(1,1)} \nn
		&\quad + \frac{1}{(16\pi^2)^2} \left[ 4 L_i^{(2,1)} - \sum_j 2 L_j^{(1,0)} \frac{\p L_i^{(1,1)}}{\p L_j^r} - \sum_j 2 L_j^{(1,1)} \frac{\p L_i^{(1,0)}}{\p L_j^r} - \sum_j 2 K_j^{(1,1)} \frac{\p L_i^{(1,0)}}{\p K_j^r} \right] \nn
		&\quad + \frac{1}{\varepsilon} \frac{1}{(16\pi^2)^2} \left[ 4 L_i^{(2,2)} - \sum_j 2 L_j^{(1,1)} \frac{\p L_i^{(1,1)}}{\p L_j^r} \right]  + \O(\text{3-loop}) + \O(\varepsilon) \, ,
\end{align}
where $L_i$ denote collectively mass matrices, gauge couplings, and coefficients of physical higher-dimension operators, whereas $K_i$ stand as usual for the coefficients of evanescent operators. The 't~Hooft consistency condition~\cite{tHooft:1972tcz,tHooft:1973mfk}
\begin{equation}
	L_i^{(2,2)} = \frac{1}{2} \sum_j L_j^{(1,1)} \frac{\p L_i^{(1,1)}}{\p L_j^r} \, ,
\end{equation}
implies that the RGEs of physical parameters are finite. This is necessarily the case in the physical basis with real and diagonal mass matrices. In our calculation with generic non-diagonal and non-Hermitian mass matrices, we find that after removal of the redundant operators the consistency condition is not automatically fulfilled for the counterterms to the mass matrices of quarks and charged leptons, apparently leading to divergent RGEs. This issue has been observed previously and discussed in detail in Refs.~\cite{Bednyakov:2014pia,Herren:2017uxn,Herren:2021yur,Zhang:2025ywe}.\footnote{See also Ref.~\cite{Manohar:2024xbh} for a discussion of infinite field anomalous dimensions.} It is related to the fact that removing redundant operators does not fix all parameters of the field redefinitions~\eqref{eq:FermionFieldRedefinition}, but leaves the combinations
\begin{equation}
	\label{eq:UnconstrainedChiralRotations}
	A_L^\psi - A_L^\psi{}^\dagger \, , \quad A_R^\psi - A_R^\psi{}^\dagger \, , \quad B_L^\psi - B_R^\psi{}^\dagger \, , \quad B_R^\psi - B_L^\psi{}^\dagger \, , \quad C_L^\psi - C_L^\psi{}^\dagger \, , \quad C_R^\psi - C_R^\psi{}^\dagger
\end{equation}
free, which are additional chiral transformations that can be used, e.g., to rotate to the basis of mass eigenstates~\cite{Jenkins:2009dy,Jenkins:2017dyc,Dekens:2019ept}. We find indeed that we can restore the 't~Hooft consistency condition and hence render the RGEs of the mass matrices finite by performing an additional chiral field redefinition of $\O(1/\varepsilon^2)$. This is reminiscent of the appearance of gauge-parameter dependences in finite one-loop contributions, which were found in Refs.~\cite{Dekens:2019ept,Naterop:2023dek} and can be similarly removed by chiral field redefinitions.

In the LEFT, such a violation of the consistency condition in the two-loop renormalization of the mass matrices appears already at dimension five due to the insertion of dipole operators. The mechanism at dimension six is analogous, but given the more compact expressions, we use the results of the dimension-five calculation of Ref.~\cite{Naterop:2024cfx} as illustration. Consider the case of electrons: after field redefinitions, the relevant divergent one-loop counterterms up to dimension five are given by~\cite{Jenkins:2017dyc,Naterop:2023dek}
\begin{align}
	M_e^{(1,1)} \Big|_\text{dim.5} &= -3 e^2 \q_e^2 M_e + 6 e \q_e \left( L_{e\gamma}^\dagger M_e^\dagger M_e + M_e M_e^\dagger L_{e\gamma}^\dagger  \right) \, , \nn
	L_{e\gamma}^{(1,1)} \Big|_\text{dim.5} &= - \frac{1}{2} e^2 (b^e_{0,0} - 10 \q_e^2) L_{e\gamma} \, ,
\end{align}
with the coefficient of the one-loop QED $\beta$-function $b_{0,0}^e$ defined in Eq.~\eqref{eq:OneLoopBetaFunction}. At dimension five, the two-loop double pole is given by
\begin{align}
	M_e^{(2,2)} \Big|_\text{dim.5} &= \left( \frac{9}{2} e^4 \q_e^4 - 3 e \q_e^2 e^{(1,1)} \right) M_e + \frac{3}{2} e^3 \q_e^3 (\xi+3) M_e L_{e\gamma} M_e \nn
		&\quad - \frac{3}{4} e^3 \q_e \left( 4 b^e_{0,0} + \q_e^2 (\xi+19) \right)  \left( L_{e\gamma}^\dagger M_e^\dagger M_e + M_e M_e^\dagger L_{e\gamma}^\dagger  \right) \, .
\end{align}
This leads to a violation of the consistency condition of the form
\begin{equation}
	M_e^{(2,2)} - \frac{1}{2} \sum_j L_j^{(1,1)} \frac{\p M_e^{(1,1)}}{\p L_j^r} = - \frac{3}{4} e^3 \q_e^3 (\xi+3) \left( L_{e\gamma}^\dagger M_e^\dagger M_e + M_e M_e^\dagger L_{e\gamma}^\dagger - 2 M_e L_{e\gamma} M_e \right) \, .
\end{equation}
However, after performing the additional chiral rotations\footnote{We are using the short-hand notation $\{ X \}_1 = X / (16\pi^2)$ and $\{ X \}_2 = X / (16\pi^2)^2$.}
\begin{align}
	A_L^e - A_L^e{}^\dagger &= - \left\{ \frac{3 e^3 \q_e^3 (\xi+3)}{2\varepsilon^2} \right\}_2 \left( L_{e\gamma} M_e - M_e^\dagger L_{e\gamma}^\dagger \right) \, , \nn
	A_R^e - A_R^e{}^\dagger &= - \left\{ \frac{3 e^3 \q_e^3 (\xi+3)}{2\varepsilon^2} \right\}_2 \left( L_{e\gamma}^\dagger M_e^\dagger - M_e L_{e\gamma} \right) \, ,
\end{align}
the consistency condition is restored. At dimension five, we find that these chiral rotations fulfill
\begin{equation}
	\tr[A_L - A_L^\dagger] = - \tr[A_R - A_R^\dagger] \, ,
\end{equation}
i.e., they contain an axial part, whereas for dimension-six insertions, the necessary chiral rotations are traceless.

%% file: sections/Computation.tex

\section{Computation}
\label{sec:Computation}

\subsection{Setup and checks of the computation}

For the computation of the two-loop LEFT RGEs due to dimension-six operator insertions, we make use of the same setup as described in Refs.~\cite{Naterop:2024cfx,Naterop:2025lzc}. We generate the Feynman diagrams with \texttt{QGRAF}~\cite{Nogueira:1991ex} and we use our own routines written in \texttt{Mathematica}, \texttt{FORM}~\cite{Vermaseren:2000nd,Ruijl:2017dtg}, and \texttt{Symbolica} for the application of Feynman rules, color, Dirac, and Lorentz algebra, and the infrared rearrangement in terms of the local $\bar R$-operation in combination with the introduction of an auxiliary mass~\cite{Chetyrkin:1997fm}. Two independent implementations enable cross checks of the calculation. We check the 't~Hooft consistency condition for the $1/\varepsilon^2$ divergences, after performing the necessary field redefinitions: this provides a basic check of the consistency of the two-loop calculation with the one-loop counterterms. However, the consistency condition does not provide a check of the two-loop RGEs, which depend on the $1/\varepsilon$ two-loop divergences. These terms are sensitive to scheme dependences, and errors in the infrared rearrangement would typically affect the $1/\varepsilon$ poles only. A much stronger check of the final results is provided by gauge-parameter independence. For this reason, we keep generic QED and QCD gauge parameters throughout, even if this comes at the price of higher computational cost compared to, e.g., Feynman gauge. Another powerful check is the cancellation of terms in the RGEs that violate chiral spurion symmetry.

\subsection{Tensor reduction and integration-by-parts relations}

The complexity of the calculation in the HV scheme is significant: the necessary dimensional splitting of the Dirac and Lorentz algebra into four-dimensional and evanescent contributions leads to much larger intermediate expressions compared to a similar calculation in the NDR scheme. For the computation of the two-loop diagrams, our two implementations rely on two different methods: in one of them, we compute directly the 1PI Green's functions. Evanescent loop momenta are written as contractions of the evanescent metric tensor with $D$-dimensional loop momenta. This procedure leads to two-loop tensor integrals in $D$ dimensions, which are reduced to scalar integrals using standard methods. In the second implementation, we compute the contribution of the two-loop diagrams to physical counterterms by applying a complete set of projectors for the physical sector (in the present work, we do not consider evanescent two-loop counterterms). This leads to a large number of Dirac traces, which can be efficiently pre-computed. The application of projectors effectively performs an efficient tensor reduction. 

In the NDR scheme, the application of projectors would directly lead to standard scalar two-loop integrals, which can be reduced using known integration-by-parts (IBP) relations~\cite{Chetyrkin:1997fm}. In the HV scheme, even when applying projectors we encounter the following scalar two-loop tadpole integrals after infrared rearrangement:
\begin{equation}
	I_{m_1,m_2,m_3}^{n_1,n_2,n_3}(M^2) = \int \frac{d^D\ell_1}{(2\pi)^D} \frac{d^D\ell_2}{(2\pi)^D} \frac{(\hat\ell_1^2)^{n_1} (\hat\ell_2^2)^{n_2} (\hat\ell_3^2)^{n_3}}{(\ell_1^2 - M^2)^{m_1} (\ell_2^2 - M^2)^{m_2} (\ell_3^2 - M^2)^{m_3}} \, ,
\end{equation}
where $\ell_3 = \ell_1 + \ell_2$. These integrals can be written as contractions of evanescent metric tensors with a $D$-dimensional tensor integral. Alternatively, using IBP relations
\begin{equation}
	0 = \{ g_{\mu\nu}, \hat g_{\mu\nu} \} \times \int \frac{d^D\ell_1}{(2\pi)^D} \frac{d^D\ell_2}{(2\pi)^D} \frac{\p}{\p {\ell_1}_\mu} \left( \ell_{1,2}^\nu \; \frac{\cdots}{\cdots} \right) \, ,
\end{equation}
one can derive the relation
\begin{equation}
	m_1 I_{m_1+1,m_2,m_3}^{n_1+1,n_2,n_3} + m_2 I_{m_1,m_2+1,m_3}^{n_1,n_2+1,n_3} + m_3 I_{m_1,m_2,m_3+1}^{n_1,n_2,n_3+1} = (n_1+n_2+n_3+D-4) I_{m_1,m_2,m_3}^{n_1,n_2,n_3} \, ,
\end{equation}
which allows us to reduce the integrals to the form
\begin{equation}
	I_{m_1,m_2,m_3}^{n_1,n_2}(M^2) := I_{m_1,m_2,m_3}^{n_1,n_2,0}(M^2) \, .
\end{equation}
Consider the integral over $\ell_2$:
\begin{align}
	I^{n}_{m_2,m_3}(\ell_1,M^2) &= \int \frac{d^D\ell_2}{(2\pi)^D} \frac{(\hat\ell_2^2)^{n/2}}{(\ell_2^2 - M^2)^{m_2} (\ell_3^2 - M^2)^{m_3}} \nn
		&= \hat g_{\mu_1\mu_2} \cdots \hat g_{\mu_{n-1} \mu_n} \int \frac{d^D\ell_2}{(2\pi)^D} \frac{\ell_2^{\mu_1} \cdots \ell_2^{\mu_n}}{(\ell_2^2 - M^2)^{m_2} (\ell_3^2 - M^2)^{m_3}} \nn
		&= \hat g_{\mu_1\mu_2} \cdots \hat g_{\mu_{n-1} \mu_n} \begin{aligned}[t] &\bigg[ (g^{\mu_1\mu_2} \cdots g^{\mu_{n-1}\mu_n}  + \text{perm.} ) I_{m_2,m_3}^{n,0} \\
			&+ (\ell_1^{\mu_1} \ell_1^{\mu_2} g^{\mu_3\mu_4} \cdots g^{\mu_{n-1}\mu_n}  + \text{perm.} ) I_{m_2,m_3}^{n,2} \\
			&+ \ldots \\
			&+ (\ell_1^{\mu_1} \cdots \ell_1^{\mu_n}  + \text{perm.} ) I_{m_2,m_3}^{n,n} \bigg] \, .\end{aligned}
\end{align}
The solution of this tensor decomposition allows us to relate the scalar tadpole integrals to linear combinations of integrals of the simple type
\begin{equation}
	\label{eq:TwoLoopIntegralOneEvanescentSquaredMomentum}
	I_{f}^{n}(M^2) := \int \frac{d^D\ell_1}{(2\pi)^D} \frac{d^D\ell_2}{(2\pi)^D} (\hat\ell_1^2)^{n} f(\ell_1^2,\ell_2^2,\ell_1 \cdot \ell_2,M^2) \, ,
\end{equation}
which can be reduced explicitly to standard $D$-dimensional scalar integrals as
\begin{align}
	\label{eq:TwoLoopIntegralOneEvanescentSquaredMomentumSolution}
	I_{f}^{n}(M^2) = \frac{(D-4)(D-2)}{(D+2n-4)(D+2n-2)} \int \frac{d^D\ell_1}{(2\pi)^D} \frac{d^D\ell_2}{(2\pi)^D} (\ell_1^2)^n f(\ell_1^2,\ell_2^2,\ell_1 \cdot \ell_2,M^2) \, .
\end{align}
The infrared rearrangement further generates products of two one-loop tadpole integrals, which in the HV scheme we can bring to the following form:
\begin{equation}
	J_{m_1,m_2}^{n_1,n_2,n_3;k}(M^2) = \int \frac{d^D\ell_1}{(2\pi)^D} \frac{d^D\ell_2}{(2\pi)^D} \frac{(\hat\ell_1^2)^{n_1} (\hat\ell_2^2)^{n_2} (\hat\ell_1 \cdot \hat \ell_2)^{n_3}  (\ell_1 \cdot \ell_2)^k}{(\ell_1^2 - M^2)^{m_1} (\ell_2^2 - M^2)^{m_2}} \, .
\end{equation}
IBP relations lead to
\begin{align}
	J_{m_1,m_2}^{n_1,n_2,n_3+1;k} &= \frac{2m_1}{k+1} J_{m_1+1,m_2}^{n_1+1,n_2,n_3;k+1} - \frac{2n_1 + n_3 + D - 4}{k+1} J_{m_1,m_2}^{n_1,n_2,n_3;k+1} \, , \nn
	J_{m_1+1,m_2}^{n_1,n_2,n_3;k} &= \frac{2n_1 + n_3 + D - 2m_1+k}{2 m_1 M^2} J_{m_1,m_2}^{n_1,n_2,n_3;k} \, , \nn
	J_{m_1,m_2+1}^{n_1,n_2,n_3;k} &= \frac{2n_2 + n_3 + D - 2m_2+k}{2 m_2 M^2} J_{m_1,m_2}^{n_1,n_2,n_3;k} \, ,
\end{align}
hence the products of two tadpoles can be reduced to integrals of the type
\begin{equation}
	J^{n_1,n_2;k}(M^2) := J_{1,1}^{n_1,n_2,0;k}(M^2) \, .
\end{equation}
To solve them, we use again a tensor decomposition. Consider the integral over $\ell_2$:
\begin{align}
	J^{n;k}_{m_2}(\ell_1,M^2) &= \int \frac{d^D\ell_2}{(2\pi)^D} \frac{(\hat\ell_2^2)^{n/2} (\ell_1 \cdot \ell_2)^k}{(\ell_2^2 - M^2)^{m_2}} \nn
		&= \hat g_{\mu_1\mu_2} \cdots \hat g_{\mu_{n-1} \mu_n} \int \frac{d^D\ell_2}{(2\pi)^D} \frac{\ell_2^{\mu_1} \cdots \ell_2^{\mu_n} (\ell_1 \cdot \ell_2)^k}{(\ell_2^2 - M^2)^{m_2}} \nn
		&= \hat g_{\mu_1\mu_2} \cdots \hat g_{\mu_{n-1} \mu_n} \begin{aligned}[t] &\bigg[ (g^{\mu_1\mu_2} \cdots g^{\mu_{n-1}\mu_n}  + \text{perm.} ) J_{m_2}^{n,0;k} \\
			&+ (\ell_1^{\mu_1} \ell_1^{\mu_2} g^{\mu_3\mu_4} \cdots g^{\mu_{n-1}\mu_n}  + \text{perm.} ) J_{m_2}^{n,2;k} \\
			&+ \ldots \\
			&+ (\ell_1^{\mu_1} \cdots \ell_1^{\mu_n}  + \text{perm.} ) J_{m_2}^{n,n;k} \bigg] \, .\end{aligned}
\end{align}
As before, inserting the tensor decomposition into the integral over $\ell_1$ leads to simple integrals of the form~\eqref{eq:TwoLoopIntegralOneEvanescentSquaredMomentum} with solution~\eqref{eq:TwoLoopIntegralOneEvanescentSquaredMomentumSolution}.

%% file: sections/Results.tex

\section{Results}
\label{sec:Results}

\subsection{Scheme dependence and structure of the RGEs}

Due to the presence of the symmetry-restoring finite one-loop counterterms derived in Ref.~\cite{Naterop:2023dek}, our results for the two-loop RGEs respect chiral spurion symmetry. However, similarly to the situation at dimension five~\cite{Naterop:2024cfx}, this is not automatically the case for the RGEs of the mass matrices of charged leptons and quarks: we find that after removing redundant operators, an additional chiral field redefinition in terms of $A_L - A_L^\dagger$ and $A_R - A_R^\dagger$ is necessary to bring the RGEs for the mass matrices into a chirally symmetric form in the spurion sense. While at dimension five, these field redefinitions contained an axial part, for dimension-six operator insertions we find instead
\begin{equation}
	\tr[ A_L - A_L^\dagger ] = \tr[ A_R - A_R^\dagger ] \, .
\end{equation}
Since these field redefinitions are of $\O(1/v^2)$ in the LEFT power counting, at dimension six they do not affect any results beyond the mass matrices.

The complete set of two-loop RGEs for the LEFT due to insertions of dimension-six operators is provided as supplementary material in the form of a \texttt{Mathematica} notebook. We use the same conventions as in Ref.~\cite{Naterop:2023dek}, which for convenience we reproduce in App.~\ref{sec:ConventionsSupplement}. As in Ref.~\cite{Naterop:2024cfx}, we write the RGEs in the form
\begin{equation}
	\label{eq:RGENotation}
	\dot X = \frac{d}{d\log\mu} X = \frac{1}{16\pi^2} [\dot X]_1 + \frac{1}{(16\pi^2)^2} [\dot X]_2 \, .
\end{equation}
The scheme-independent one-loop contribution to the RGEs $[\dot X]_1$ have been computed in Ref.~\cite{Jenkins:2017dyc}. In the supplementary material, we provide the two-loop contribution $[\dot X]_2$ to the RGEs due to the insertion of dimension-six operators in the HV scheme.

Some peculiar results can be observed: for example, we obtain the following RGE for the neutrino--down-quark vector operators for different quark flavors
\begin{align}
	\label{eq:RGEsVectorNuD}
	\bigg[ \dlwc{\nu d}{LL}[V][prst] \bigg]_2 &= 4 \left( e^4 \q_d^2 b_{0,0}^e + g^4 C_F b_{0,0}^g \right) \lwc{\nu d}{LL}[V][prst] \, , \quad \text{for } s \neq t \, , \nn
	\bigg[ \dlwc{\nu d}{LR}[V][prst] \bigg]_2 &= 4 \left( e^4 \q_d^2 b_{0,0}^e + g^4 C_F b_{0,0}^g \right) \lwc{\nu d}{LR}[V][prst] \, , \quad \text{for } s \neq t \, ,
\end{align}
consistent with Ref.~\cite{Buras:1989xd}. As the vector current is a conserved current, one might expect that these operators should have vanishing anomalous dimension. This is indeed observed in NDR~\cite{Aebischer:2025hsx}. However, the non-vanishing two-loop anomalous dimensions of vector currents in different schemes are in fact well known~\cite{Altarelli:1980fi,Buras:1989xd}. Our scheme includes finite symmetry-restoring counterterms, given by~\cite{Naterop:2023dek}
\begin{align}
	\label{eq:FiniteCountertermsLVLXnud}
	\delta_\text{fin}^\chi\bigg( \lwc{\nu d}{LL}[V][prst] \bigg) &= 2 \left\{ e^2 \q_d^2 + g^2 C_F \right\}_1 \lwc{\nu d}{LR}[V][prst] \, , \nn
	\delta_\text{fin}^\chi\bigg( \lwc{\nu d}{LR}[V][prst] \bigg) &= 2 \left\{ e^2 \q_d^2 + g^2 C_F \right\}_1 \lwc{\nu d}{LL}[V][prst] \, .
\end{align}
These finite renormalizations are minimal in the chiral basis, i.e., we do not include finite symmetry-preserving counterterms: in other words, our finite {\em symmetry-restoring} counterterms consist of purely {\em symmetry-breaking} contributions, which compensate the symmetry-breaking effects from the one-loop diagrams. The consequence of this scheme choice is that even in the absence of an axial-vector component, i.e., for $\lwc{\nu d}{LL}[V] = \lwc{\nu d}{LR}[V]$, a finite renormalization is applied. This has the effect that our renormalized operators in the HV scheme are not in direct correspondence to the properly normalized conserved currents. Instead, this relation involves another finite renormalization. As usual, the scheme choice is of no physical consequence and drops out in relations between observables.\footnote{It is not unusual that the relation between renormalized operators and conserved quantities requires finite renormalizations. Due to penguin graphs, even the usual \msbar{} electromagnetic current has a non-vanishing anomalous dimension and the relation to the physical charge operator involves a finite renormalization~\cite{Collins:2005nj}.} As discussed for the baryon-number-violating sector in Ref.~\cite{Naterop:2025lzc}, we could modify the HV renormalization scheme by including additional symmetry-preserving finite counterterms, so that the finite renormalizations vanish in the absence of $P$-odd interactions. In the present example, we could use modified finite counterterms
\begin{align}
	\widetilde\delta_\text{fin}^\chi\bigg( \lwc{\nu d}{LL}[V][prst] \bigg) &= 2 \left\{ e^2 \q_d^2 + g^2 C_F \right\}_1 \left( \lwc{\nu d}{LR}[V][prst] - \lwc{\nu d}{LL}[V][prst] \right) \, , \nn
	\widetilde\delta_\text{fin}^\chi\bigg( \lwc{\nu d}{LR}[V][prst] \bigg) &= 2 \left\{ e^2 \q_d^2 + g^2 C_F \right\}_1 \left( \lwc{\nu d}{LL}[V][prst] - \lwc{\nu d}{LR}[V][prst] \right) \, ,
\end{align}
which differ from Eq.~\eqref{eq:FiniteCountertermsLVLXnud} only by symmetry-preserving contributions. In this modified scheme, the two-loop anomalous dimensions~\eqref{eq:RGEsVectorNuD} would vanish.

The RGEs for the four-fermion LEFT operators were recently obtained in the NDR scheme~\cite{Aebischer:2025hsx}. Due to the scheme dependences, a direct comparison of our results with the NDR results is not possible. A dedicated analysis of the scheme translation, as we have done it in Ref.~\cite{Naterop:2025lzc} for the baryon-number-violating sector, is left for future work.

In contrast to an NDR calculation, we emphasize that the HV scheme does not suffer from algebraic inconsistencies, hence it allows us to directly obtain all the two-loop RGEs, whereas the NDR results of Ref.~\cite{Aebischer:2025hsx} required some detours using one-loop basis changes in order to avoid ill-defined $\gamma_5$-odd traces. The HV scheme allows us to obtain not only the four-fermion anomalous dimensions, but the complete dimension-six RGEs including down-mixing into lower-dimension operators in the presence of masses. These effects cannot be obtained in pure NDR without an additional prescription (e.g., for a reading point of fermion traces~\cite{Kreimer:1989ke,Korner:1991sx}), since ill-defined $\gamma_5$-odd traces show up already in finite one-loop contributions.

The mixing structure of the LEFT RGEs is restricted by the fact that the gauge interactions are flavor diagonal. Together with chiral (spurion) symmetry, this allows one to split the set of Wilson coefficients into sectors that remain invariant under RG-evolution, as has been discussed in detail in Ref.~\cite{Aebischer:2025hsx}. Our results in the HV scheme respect the general mixing pattern, due to the restoration of chiral symmetry by finite counterterms. In the presence of masses, the down-mixing exhibits a rich structure: since mass insertions lead to a chirality flip, a large set of four-fermion operators with scalar, vector, and tensor Dirac structures mix into the mass matrices or dipole operators.

\subsection{Mixing and renormalization of gauge couplings and three-gluon operators}

Certain mixings are absent at two loops although they cannot be excluded by symmetry arguments. One example are mixings of four-fermion operators into the gauge couplings, theta parameters, and three-gluon operator coefficients. The reason is that at two loops, an insertion of a four-fermion operator into a pure gauge-boson Green's function necessarily leads to topologies that factorize into two one-loop graphs. In our scheme, these diagrams only lead to $1/\varepsilon^2$ divergences and do not generate single poles in $\varepsilon$, hence they do not contribute to the RGEs~\cite{Jenkins:2023rtg,Naterop:2025factorizable}.
This leads to simple expressions for the two-loop RGEs of these parameters. The two-loop running of the gauge couplings due to the insertion of dimension-six operators is
\begin{align}
	\left[ \dot e \right]_2 &= 0 \, , \nn
	\left[ \dot g \right]_2 &= 134 g^4 N_c \left( \< M_u M_u^\dagger \> + \< M_d M_d^\dagger \> \right)  L_G \, ,
\end{align}
whereas the running of the $\theta$-parameters is given by
\begin{align}
	\left[ \dot \theta_\mathrm{QED} \right]_2 &= 0 \, , \nn
	\left[ \dot \theta_\mathrm{QCD} \right]_2 &= 2144 \pi^2 g N_c \left( \< M_u M_u^\dagger \> + \< M_d M_d^\dagger \> \right)  L_{\widetilde G} \, .
\end{align}
We observe that these contributions to the gauge couplings fulfill holomorphy with respect to the three-gluon operator coefficients, i.e., the linear combination
\begin{equation}
	\tau_\mathrm{QCD} = i \frac{4\pi}{g^2} + \frac{\theta_\mathrm{QCD}}{2\pi}
\end{equation}
only depends on the self-dual three-gluon-operator coefficient $L_G + i L_{\widetilde G}$, but not on the anti-self-dual coefficient $L_G - i L_{\widetilde G}$ (the dependence on the masses however is not holomorphic as dictated by chiral spurion symmetry). Holomorphy is also respected in the mixing of the three-gluon operators into dipole operators. We obtain
\begin{align}
	\left[ \dot L_{q\gamma} \right]_2 &= - 8 e \q_q g^3 N_c C_F M_q^\dagger (L_G + i L_{\widetilde G}) + \ldots \, , \nn
	\left[ \dot L_{qG} \right]_2 &= - \frac{1}{2} g^2 \left( 22 e^2 \q_q^2 N_c + g^2 (6 N_c b_{0,0}^g - 50 N_c^2 - 19) \right) M_q^\dagger (L_G + i L_{\widetilde G}) + \ldots \, ,
\end{align}
where the ellipses stand for omitted four-fermion contributions. We find that for the three-gluon-operator mixing into the masses, holomorphy is only respected after we perform another anti-Hermitian chiral field redefinition of the form
\begin{equation}
	A_L^q - A_L^q{}^\dagger = - \left\{ \frac{19 i g^3 N_c C_F L_{\widetilde G}}{4\varepsilon} \right\}_2 M_q^\dagger M_q \, , \quad
	A_R^q - A_R^q{}^\dagger = \left\{ \frac{19 i g^3 N_c C_F L_{\widetilde G}}{4\varepsilon} \right\}_2 M_q M_q^\dagger \, ,
\end{equation}
which contains an axial part and will impact the theta terms at the three-loop level. After this chiral rotation, the mixing of the three-gluon operators into the quark-mass matrices is
\begin{align}
	\left[ \dot M_q \right]_2 &= 13 g^3 N_c C_F M_q M_q^\dagger M_q (L_G - i L_{\widetilde G}) + \ldots \, .
\end{align}

The two-loop RGEs of the three-gluon operators are
\begin{align}
	\left[ \dot L_G \right]_2 &= \left( 6 e^2 g^2 (n_u \q_u^2 + n_d \q_d^2) + n_q \frac{g^4 \left(N_c^2-9\right)}{3N_c} + \frac{47 g^4 N_c^2}{3}\right) L_G \, , \nn*
	\left[ \dot L_{\widetilde G} \right]_2 &= \left( 6 e^2 g^2 (n_u \q_u^2 + n_d \q_d^2) + n_q \frac{g^4 \left(N_c^2-9\right)}{3N_c} + \frac{47 g^4 N_c^2}{3}\right) L_{\widetilde G} \, ,
\end{align}
where $n_q = n_u + n_d$.

Finally, we remark that we find a mixing of the $CP$-even three-gluon operator at two loops into vector-type four-fermion operators, both semi-leptonic ones and four-quark operators. For the explicit results, we refer to the supplementary material. A mixing of three-gluon operators into scalar- and tensor-type four-fermion operators is excluded at any loop order, as no chirality flip can happen at dimension six. In the case of the $CP$-odd three-gluon operator, there is no mixing at all into four-fermion operators: this is prohibited, because a mixing could only occur with flavor- and chirality-conserving $CP$-odd four-fermion operators, which do not exist~\cite{Khatsimovsky:1987fr,Buhler:2023gsg}.

Our RGE results for the three-gluon-operator insertions can be compared to existing partial results in the literature. The mixing of the $CP$-even three-gluon operator into the gauge coupling and quark mass has been computed in Ref.~\cite{Duhr:2025zqw}. There, the computation was performed in the \msbar{} scheme with a $D$-dimensional three-gluon operator
\begin{equation}
	\label{eq:CPE3GOprime}
	\O_{G}^\prime = f^{ABC} G_\mu^{A\nu} G_\nu^{B\rho} G_\rho^{C\mu}  \, ,
\end{equation}
which is a natural choice for an NDR calculation. In contrast, our three-gluon operator (as all physical higher-dimension operators in our basis) is defined with indices running over four space-time dimensions only,
\begin{equation}
	\label{eq:CPE3GO}
	\O_{G} = f^{ABC} \roverline{ G_\mu^{A\nu} G_\nu^{B\rho} G_\rho^{C\mu} }  \, ,
\end{equation}
which is convenient as the HV scheme requires a dimensional splitting. The different operators are related by
\begin{equation}
	\O_{G}^\prime =  \O_G + 3 \E_{G1} - 3 \E_{G2} + \E_{G3} \, ,
\end{equation}
with the evanescent operators~\cite{Naterop:2023dek}
\begin{equation}
	\E_{G1} = \hat g^{\rho\sigma} f^{ABC} G_\mu^{A\nu} G_{\nu\sigma}^{B} G_\rho^{C\mu} \, , \quad
	\E_{G2} = f^{ABC} G_\mu^{A\nu} G_\nu^{B\rho} \hat G_\rho^{C\mu} \, , \quad
	\E_{G3} = f^{ABC} \hat G_\mu^{A\nu} \hat G_\nu^{B\rho} \hat G_\rho^{C\mu} \, .
\end{equation}
Because of the different scheme definitions, the mixings obtained in Ref.~\cite{Duhr:2025zqw} cannot be compared directly to our results. In App.~\ref{sec:BasisChangeCPEven}, we describe the scheme change in detail. Applying this conversion and setting $M_q = M_q^\dagger$ to avoid the appearance of $\gamma_5$, we find for the three-gluon mixing into the gauge coupling and the quark masses
\begin{align}
	\label{eq:LGmixingsNDR}
	\left[ \dot g \right]_2^\mathrm{NDR} &= 138 g^4 N_c \left( \<M_u^2\> + \<M_d^2\> \right) L_G' + \ldots \, , \nn
	\left[ \dot M_q \right]_2^\mathrm{NDR} &= - \frac{7}{2} g^3 N_c C_F M_q^3 L_G' + \ldots \, ,
\end{align}
which agrees with Ref.~\cite{Duhr:2025zqw}. Similarly, we also reconstruct the NDR result for the two-loop running of the $CP$-even three-gluon operator,
\begin{align}
	\label{eq:LGrunningNDR}
	\left[ \dot L_G' \right]_2^\mathrm{NDR} &= \left( 6 e^2 g^2 (n_u \q_u^2 + n_d \q_d^2 ) + n_q \frac{g^4 (5 N_c^2 - 9)}{3N_c} + \frac{25 g^4 N_c^2}{3} \right) L_G' \, ,
\end{align}
where the purely bosonic ($n_q$-independent) part agrees with Ref.~\cite{Born:2024mgz} and the $n_q$-dependence is (to our knowledge) a new result.

The renormalization of the $CP$-odd three-gluon operator has been computed in Ref.~\cite{deVries:2019nsu} to two loops in QCD and even to three loops in pure Yang--Mills theory. In the $CP$-odd sector, $\gamma_5$ and the Levi-Civita symbol appear, which are problematic in NDR. In Ref.~\cite{deVries:2019nsu}, both the HV scheme and the Larin scheme~\cite{Larin:1993tq} were applied, leading to identical results. However, the HV scheme of Ref.~\cite{deVries:2019nsu} does not correspond exactly to our scheme. The main difference again concerns the definition of the physical operator. We are using a definition with Lorentz indices running over four space-time dimensions,
\begin{equation}
	\label{eq:CP3GO}
	\O_{\widetilde G} = f^{ABC} \roverline{ \widetilde G_\mu^{A\nu} G_\nu^{B\rho} G_\rho^{C\mu} }  \, ,
\end{equation}
whereas in Ref.~\cite{deVries:2019nsu}, the range of the indices is not restricted. This leads to the relation
\begin{equation}
	\label{eq:CP3GOprime}
	\O_{\widetilde G}^\prime = f^{ABC} \widetilde G_\mu^{A\nu} G_\nu^{B\rho} G_\rho^{C\mu} = \O_{\widetilde G} + \E_{\widetilde G}  \, ,
\end{equation}
with the evanescent operator
\begin{equation}
	\E_{\widetilde G} = \hat g^{\rho\sigma} f^{ABC} \widetilde G_\mu^{A\nu} G_{\nu\sigma}^{B} G_{\rho}^{C\mu}  \, .
\end{equation}
We recomputed the two-loop RGE with this alternative operator definition, see App.~\ref{sec:BasisChangeCPOdd} for details. In this case, we reproduce the two-loop QCD results of Ref.~\cite{deVries:2019nsu} if we employ strict \msbar{}, as done in that reference. However, in contrast to the $CP$-even three-gluon operator the use of \msbar{} is problematic for the insertion of the $CP$-odd counterpart. The reason is that evanescent two-quark operators are generated at dimension six, due to the fact that the Lorentz indices contracted with the Levi-Civita symbol are restricted to four space-time dimensions. Some of the one-loop-generated evanescent operators mix back into the physical sector if minimal subtraction is used. Neglecting this evanescent mixing when using \msbar{}, as done in Ref.~\cite{deVries:2019nsu}, leads to incorrect results, i.e., not all next-to-leading logarithms are captured by the diagonal RGE. The standard procedure is not to include the mixing of the evanescent sector into the physical sector in \msbar{}, but rather to avoid such a mixing by renormalizing the evanescent sector, which implies a departure from minimal subtraction~\cite{Dugan:1990df,Herrlich:1994kh}. Applying such a finite renormalization with the operator definition~\eqref{eq:CP3GOprime}, we find that the $n_q$-dependent part of the two-loop RGE is different from the strict \msbar{} result of Ref.~\cite{deVries:2019nsu}. Instead, we find
\begin{align}
	\label{eq:CP3GORGEprime}
	\left[ \dot L_{\widetilde G}' \right]_2 &= \left( 6 e^2 g^2 (n_u \q_u^2 + n_d \q_d^2) + n_q \frac{g^4 \left(7N_c^2-27\right)}{9N_c} + \frac{119 g^4 N_c^2}{9}\right) L_{\widetilde G}' \, .
\end{align}
We note that these scheme definitions are relevant not only for the two-loop RGE, but also for one-loop matching calculations of the three-gluon operator~\cite{Cirigliano:2020msr,Crosas:2023anw}.

%% file: sections/Conclusions.tex

\section{Conclusions}
\label{sec:Conclusions}

In this article, we have presented the complete two-loop RGEs of the LEFT due to the insertion of baryon-number-conserving dimension-six operators. While the one-loop RGEs are scheme independent~\cite{Jenkins:2017dyc}, at two loops scheme dependences are present. Our results are provided in the algebraically consistent HV scheme defined in Ref.~\cite{Naterop:2023dek}, which enables in particular a rigorous approach to treat the $CP$-odd sector of the theory. Finite one-loop renormalizations ensure a decoupling of the evanescent sector and restore global chiral spurion symmetry. Consequently, the two-loop RGEs of physical operators are independent of the coefficients of evanescent operators and they respect chiral spurion symmetry.
The results obtained here should either be combined with one-loop matching and matrix-element calculations in the same scheme, or, if a different scheme is used, the scheme translations need to be taken into account. As an example, we studied the translation to the NDR scheme for the $CP$-even three-gluon operator, a case where $\gamma_5$ does not appear. Systematically establishing this contact to the more widely used but algebraically inconsistent NDR scheme, similarly to Ref.~\cite{DiNoi:2025uan} in the case of SMEFT, is left for future work. Our results are a further step towards obtaining the complete two-loop RGEs of the LEFT at dimension six: the last missing part is the contribution of double insertions of dimension-five operators, which will be presented in a forthcoming publication~\cite{Naterop:2025inPrep}.

%% file: sections/ConventionsSupplement.tex

\section{Conventions for the supplementary material}
\label{sec:ConventionsSupplement}

\begin{table}[t]
	\centering
	\small
	\mbox{} \\[-0.5cm]
	\begin{tabular}{lllc}
		\toprule
		variable							& code name				& explanation						\\
		\midrule
		\midrule
		$N_c$							& \verb$Nc$				& number of colors \\
		$C_F$							& \verb$CF$				& $SU(3)_c$ fundamental Casimir invariant \\
		\midrule
		$n_e$							& \verb$nf[e]$				& number of charged lepton flavors \\
		$n_u$							& \verb$nf[u]$				& number of up-type quark flavors \\
		$n_d$							& \verb$nf[d]$				& number of down-type quark flavors \\
		\midrule
		$b_{0,0}^e$						& \verb$b00e$				& coefficient of QED $\beta$-function \\
		$b_{0,0}^g$						& \verb$b00g$				& coefficient of QCD $\beta$-function \\
		\midrule
		$\q_e = -1$						& \verb$q[e]$				& electron charge \\
		$\q_u = 2/3$						& \verb$q[u]$				& up-quark charge \\
		$\q_d = -1/3$						& \verb$q[d]$				& down-quark charge \\
		\midrule
		$e$, $g$							& \verb$e$, \verb$g$				& QED and QCD gauge couplings \\
		$\theta_\mathrm{QED}$, $\theta_\mathrm{QCD}$		& \verb$[\Theta]QED$, \verb$[\Theta]QCD$ 		& QED and QCD theta parameters \\
		\midrule
		$M_\nu, M_\nu^\dagger$				& \verb$M[nu]$, \verb$Mdag[nu]$	& neutrino mass matrix \\
		$M_e, M_e^\dagger$				& \verb$M[e]$, \verb$Mdag[e]$		& charged-lepton mass matrix \\
		$M_u, M_u^\dagger$				& \verb$M[u]$, \verb$Mdag[u]$		& up-quark mass matrix \\
		$M_d, M_d^\dagger$				& \verb$M[d]$, \verb$Mdag[d]$		& down-quark mass matrix \\
		\midrule
		$\delta_{pr}$						& \verb$kd[f,p,r]$			& flavor Kronecker delta for fermion type $f$ \\
		$\tr[A \cdots B]$					& \verb$flTr[A,...,B]$			& trace in flavor space \\
		$(A \cdots B)_{pr}$					& \verb$FCHN[{A,...,B},p,r]$	& flavor chain: element $p,r$ of a product \\
																&&  of flavor-space matrices \\
		\midrule
		$\lwc{uu}{LR}[V8][]$					& \verb$LV8LRuu$			& Wilson coefficients \\
		\ldots \\
		\midrule
		$[\dot X ]_2$						& \verb$dot2[X]$			& two-loop contribution to RGE, see Eq.~\eqref{eq:RGENotation} \\
		\bottomrule
	\end{tabular}
	\caption{LEFT variables appearing in the code with the two-loop results, provided as supplementary material.}
	\label{tab:CodeVariables}
\end{table}

The complete results for the two-loop RGEs in the LEFT due to insertions of dimension-six operators are provided as supplementary material in the form of a \texttt{Mathematica} notebook. In addition to the baryon-number-conserving operators, we also include the results for the baryon-number-violating operators derived in Ref.~\cite{Naterop:2025lzc}. In most cases, we write the dependence on the number of flavors in terms of the coefficients of the one-loop QED and QCD beta functions
\begin{equation}
	\label{eq:OneLoopBetaFunction}
	b^e_{0,0} = -\frac{4}{3} \left( n_e \q_e^2 + N_c (n_u \q_u^2 + n_d \q_d^2 ) \right) \, , \quad
	b^g_{0,0} = \frac{11}{3} N_c - \frac{2}{3} ( n_u + n_d ) \, .
\end{equation}
In some parts of our results, we use the coefficient of the quadratic Casimir $C_F = (N_c^2-1)/(2N_c)$ of $SU(N_c)$ for compact notation, but we do not claim them to be valid for different gauge groups. In particular, we do not distinguish $C_A$ from $N_c$. In the case of baryon-number-violating operators, one has to use $N_c = 3$~\cite{Naterop:2025lzc}. 

The results are written in the form of replacement rules for the two-loop contribution to the RGEs, using the same conventions as Ref.~\cite{Naterop:2023dek}. The symbols appearing in the text files are listed in Table~\ref{tab:CodeVariables}. We are using a matrix-style notation in flavor space, so that the only indices are the open indices of the replacement rule. This notation is extended to rank-4 flavor tensors as
\begin{align}
	\mathrm{FFA}(\lwc{ee}{LR}[V])_{pr} \, \mathrm{FFB}(\lwc{ee}{LR}[V])_{st} := \lwc{ee}{LR}[V][prst] \, ,
\end{align}
where the first two flavor indices are attached to the symbol $\mathrm{FFA}$ and the last two indices to the symbol $\mathrm{FFB}$, which always need to appear together in an expression.
As an example, the contribution
\begin{minipage}{\linewidth}
\mbox{}
\begin{lstlisting}[frame=trBL,basicstyle=\small\ttfamily]
 dot2[L\[Nu]\[Gamma][fm2_, fm1_]] ->
	-72*e^3*q[e]^3*FCHN[{FFA[LTLL\[Nu]e]}, fm2, fm1]*flTr[FFB[LTLL\[Nu]e], Mdag[e]]
	+ ...
\end{lstlisting}
\end{minipage}
corresponds to
\begin{equation}
	\Big[ \dlwc{\nu\gamma}{}[][pr] \Big]_2 =  - 72 e^3 \q_e^3 \lwc{\nu e}{LL}[T][prvw] \left[ M_e^\dagger \right]_{wv} + \ldots
\end{equation}
in the index notation of Ref.~\cite{Jenkins:2017dyc}. For the Wilson coefficients of Hermitian conjugate operators, we follow the convention of Refs.~\cite{Jenkins:2017dyc,Dekens:2019ept}, e.g.,
\begin{align}
	\lwc{e\gamma}{\dagger}[][pr] &:= \Big(\lwc{e\gamma}{}[][rp]\Big)^* \, , \quad
	\lwc{uddu}{RR\dagger}[S1][prst] := \Big(\lwc{uddu}{RR}[S1][rpts]\Big)^* \, .
\end{align}
The RGEs for the Wilson coefficients of Hermitian conjugate operators are trivially related and not listed explicitly.

%% file: sections/BasisChange.tex

\section{Basis change for three-gluon operators}
\label{sec:BasisChange}

\subsection[$CP$-even three-gluon operator]{\boldmath $CP$-even three-gluon operator}
\label{sec:BasisChangeCPEven}

In this appendix, we derive the two-loop mixing of the $CP$-even three-gluon operator into the gauge coupling and quark mass for the alternative operator definition given in Eq.~\eqref{eq:CPE3GOprime}, which is used in Ref.~\cite{Duhr:2025zqw} using the NDR scheme. Although our calculation is in the HV scheme, for the insertions of the $CP$-even three-gluon operator we can derive the NDR results by setting $M_q = M_q^\dagger$, which avoids any appearance of $\gamma_5$.

For notational simplicity, in this appendix we use the notation
\begin{equation}
	\E_X := 3 \E_{G1} - 3 \E_{G2} + \E_{G3} \, .
\end{equation}
We relate the dimension-six operators in the primed basis (with $D$-dimensional Lorentz indices in physical operators) to our operators (with Lorentz indices in physical operators restricted to four dimensions) as
\begin{equation}
	\label{eq:CPE3GluonPrimedRelation}
	\L' = L_{G}' \O_{G}' + K_X' \E_X' + \ldots = L_{G}' \O_{G} + (K_X'  + L_{G}'  )\E_X + \ldots  \, .
\end{equation}
Since in the primed basis all interactions are defined in $D$ dimensions, no evanescent terms are generated from the one-loop insertions of the three-gluon operator $\O_G'$. Therefore, the last term of the second line in Eq.~\eqref{eq:RGEMasterFormula} is absent in the primed basis for these insertions, i.e.,
\begin{equation}
	\frac{\p K^{\prime(1,1)}_j}{\p L_{G}'} = 0 \, .
\end{equation}
In NDR, there are no chiral-symmetry-breaking counterterms, hence the two-loop RGEs due to insertions of $\O_G'$ directly derive from
\begin{equation}
	\frac{\p}{\p L_G'} [ \dot L_i' ]_2 = 4 \frac{\p L_i^{\prime(2,1)}}{\p L_G'}
\end{equation}
for any parameter $L_i'$ in the primed scheme. Due to the relation~\eqref{eq:CPE3GluonPrimedRelation}, the two-loop $1/\varepsilon$ counterterm in the primed basis can be obtained from
\begin{align}
	\label{eq:CPevenTwoLoopCTPrimedBasis}
	\frac{\p L_i^{\prime(2,1)}}{\p L_{G}'} &= \frac{\p L_i^{(2,1)}}{\p L_{G}} - \sum_j \frac{\p L_{j,\chi}^{(1,0)}}{\p L_G}\frac{\p L_i^{(1,1)}}{\p L_j} + \frac{\p L_i^{(2,1)}}{\p K_{X}} - \sum_j \frac{\p L^{(1,0)}_{j,\text{ev}}}{\p K_{X}}\frac{\p L_i^{(1,1)}}{\p L_j} \, .
\end{align}
The first term on the right-hand side is the two-loop $1/\varepsilon$ counterterm in the original basis. The second term removes the effect of the finite chiral-symmetry-restoring renormalization, which is present in our HV scheme but must not be applied in NDR. The third term arises from the two-loop insertion of the additional evanescent part of $\O_{G}'$, whereas the last term corrects for the one-loop counterterm diagram with a finite renormalization of the evanescent operator, which should not be carried over to the insertion of the physical operator in the primed basis.

The dependence of two-loop counterterms on evanescent insertions satisfies a consistency condition~\cite{Naterop:2023dek}
\begin{align}
	\label{eq:EvanescentConsistencyCondition}
	\frac{\p L_i^{(2,1)}}{\p K_k^r} = \frac{1}{2} \Biggl[ &\sum_j \frac{\p L_j^{(1,0)}}{\p K_k^r} \frac{\p L_i^{(1,1)}}{\p L_j^r} + \sum_j L_j^{(1,1)} \frac{\p^2 L_i^{(1,0)}}{\p K_k^r \p L_j^r} \nn*
		&+ \sum_j \frac{\p K_j^{(1,1)}}{\p K_k^r} \frac{\p L_i^{(1,0)}}{\p K_j^r} + \sum_j K_j^{(1,1)} \frac{\p^2 L_i^{(1,0)}}{\p K_k^r \p K_j^r} \Biggr] \, ,
\end{align}
where the sums in the first line extend over physical parameters including gauge couplings, while the ones in the second line run over all evanescent coefficients $K_j$. This allows us to compute the scheme change solely using finite and divergent one-loop counterterms, leading to the results in Eqs.~\eqref{eq:LGmixingsNDR} and~\eqref{eq:LGrunningNDR}. We have verified Eq.~\eqref{eq:LGrunningNDR} as well by explicit computation of the two-loop insertion of the $D$-dimensional three-gluon operator $\O_G'$.

\subsection[$CP$-odd three-gluon operator]{\boldmath $CP$-odd three-gluon operator}
\label{sec:BasisChangeCPOdd}

In the following, we derive the two-loop RGE for the alternative definition of the $CP$-odd three-gluon operator given in Eq.~\eqref{eq:CP3GOprime}, as used in Ref.~\cite{deVries:2019nsu}, similarly to the $CP$-even case in Sect.~\ref{sec:BasisChangeCPEven} but using the HV scheme.

We relate the operators in the two bases as
\begin{equation}
	\L' = L_{\widetilde G}' \O_{\widetilde G}' + K_{\widetilde G}' \E_{\widetilde G}' + \ldots = L_{\widetilde G}' \O_{\widetilde G} + (K_{\widetilde G}'  + L_{\widetilde G}'  )\E_{\widetilde G} + \ldots  \, .
\end{equation}
The divergent physical one-loop counterterms are scheme independent, $L^{\prime(1,1)}_i = L^{(1,1)}_i$. Also the evanescent-compensating finite one-loop renormalizations of the physical operators are identical in the two bases if the evanescent operators are the same, $L^{\prime(1,0)}_{i,\text{ev}} = L^{(1,0)}_{i,\text{ev}}{}$. The symmetry-restoring finite one-loop counterterms of the physical operators differ by
\begin{equation}
	L^{\prime(1,0)}_{i,\chi}(L_{\widetilde G}',0) = L^{(1,0)}_{i,\chi}(L_{\widetilde G}',0) + L^{(1,0)}_{i,\text{ev}}(0,K_{\widetilde G}' \to L_{\widetilde G}') \Big|_{\slashed\chi} \, ,
\end{equation}
where the last term denotes the symmetry-breaking parts of the $\E_{\widetilde G}$-compensating finite renormalizations, with the coefficient $K'_{\widetilde G}$ replaced by $L'_{\widetilde G}$. These shifts only affect gluonic dipole-operator coefficients and are not relevant for the running of $L_{\widetilde G}$. Finally, most interesting at one loop is the change in the evanescent divergences, which are shifted as
\begin{align}
	K^{\prime(1,1)}_{\widetilde G}(L_{\widetilde G}',K_{\widetilde G}') &= K^{(1,1)}_{\widetilde G}(L_{\widetilde G}',K_{\widetilde G}'+L_{\widetilde G}') - L^{(1,1)}_{\widetilde G}(L_{\widetilde G}',0) \, , \nn
	K^{\prime(1,1)}_{i}(L_{\widetilde G}',K_{\widetilde G}') &= K^{(1,1)}_{i}(L_{\widetilde G}',K_{\widetilde G}'+L_{\widetilde G}') \quad \text{for } i \neq \widetilde G \, .
\end{align}
Explicitly, with
\begin{align}
	L^{(1,1)}_{\widetilde G} &= \frac{1}{2} g^2 (2 n_q + N_c) L_{\widetilde G} \, , \nn
	K^{(1,1)}_{\widetilde G} &= - g^2 N_c L_{\widetilde G} + \frac{1}{2} g^2 ( 2 n_q + 3 N_c ) K_{\widetilde G} \, ,
\end{align}
one finds
\begin{equation}
	\label{eq:K11prime}
	K^{\prime(1,1)}_{\widetilde G} = \frac{1}{2} g^2 ( 2 n_q + 3 N_c ) K'_{\widetilde G} \, ,
\end{equation}
i.e., the one-loop insertion of $\O'_{\widetilde G}$ into a three-gluon Green's function does not generate an evanescent counterterm and the $K_{\widetilde G}$ contribution to the last term of the second line in Eq.~\eqref{eq:RGEMasterFormula} is absent in the primed basis. This is not true for all evanescent coefficients: in particular, evanescent counterterms are generated from the one-loop insertion of $\O'_{\widetilde G}$ into quark--anti-quark--gluon Green's functions, as we will discuss below. 

Similarly to the case of the $CP$-even operator~\eqref{eq:CPevenTwoLoopCTPrimedBasis}, the two-loop $1/\varepsilon$ counterterm in the primed basis is obtained from\footnote{In contrast to the quark mass or gauge coupling in Eq.~\eqref{eq:CPevenTwoLoopCTPrimedBasis}, for the dimension-six three-gluon operators there is no correction due to symmetry-restoring counterterms, which vanish.}
\begin{equation}
	\frac{\p L^{\prime(2,1)}_{\widetilde G}}{\p L_{\widetilde G}'} = \frac{\p L^{(2,1)}_{\widetilde G}}{\p L_{\widetilde G}} + \frac{\p L^{(2,1)}_{\widetilde G}}{\p K_{\widetilde G}} - \frac{\p L^{(1,0)}_{\widetilde G,\text{ev}}}{\p K_{\widetilde G}}\frac{\p L^{(1,1)}_{\widetilde G}}{\p L_{\widetilde G}} \, ,
\end{equation}
The finite renormalizations are~\cite{Naterop:2023dek}
\begin{equation}
	L^{(1,0)}_{\widetilde G,\text{ev}} = \frac{1}{3} g^2 N_c K_{\widetilde G} + \frac{1}{3} g^2 \, \Im \left( \< \kwc{uDG1}{LR} - \kwc{DuG1}{LR} \> + \< \kwc{dDG1}{LR} - \kwc{DdG1}{LR} \> \right) \, , \nn
\end{equation}
where $\< \cdot \>$ stands for the trace in flavor space. The evanescent one-loop divergences due to the insertion of $\E_{\widetilde G}$ are given by
\begin{align}
	\p \kwc{uDG1}{LR\,(1,1)}[][pr] \Big/ \p K_{\widetilde G} &= \p \kwc{dDG1}{LR\,(1,1)}[][pr] \Big/ \p K_{\widetilde G} =  \frac{i}{6} g^2 N_c \delta_{pr} \, , \nn
	\p \kwc{DuG1}{LR\,(1,1)}[][pr] \Big/ \p K_{\widetilde G} &= \p \kwc{DdG1}{LR\,(1,1)}[][pr] \Big/ \p K_{\widetilde G} =  -\frac{i}{6} g^2 N_c \delta_{pr} \, .
\end{align}
Using the consistency condition~\eqref{eq:EvanescentConsistencyCondition}, this allows us to derive
\begin{equation}
	\frac{\p L^{(2,1)}_{\widetilde G}}{\p K_{\widetilde G}} = \frac{1}{18} g^4 N_c \left( 9 n_q - 5 N_c \right) \, ,
\end{equation}
which we verified by explicit calculation of the two-loop insertion of $\E_{\widetilde G}$. This finally gives us the shift in the two-loop $1/\varepsilon$ counterterm when going to the primed basis,
\begin{equation}
	\frac{\p L^{\prime(2,1)}_{\widetilde G}}{\p L_{\widetilde G}'} - \frac{\p L^{(2,1)}_{\widetilde G}}{\p L_{\widetilde G}} = \frac{1}{18} g^4 N_c \left( 3 n_q - 8 N_c \right) \, .
\end{equation}
With that result, we have all the ingredients to calculate the two-loop RGE of $L_{\widetilde G}'$. 

The correction to the RGE in our original basis due to finite renormalizations of evanescent insertions in Eq.~\eqref{eq:RGEMasterFormula} is given by
\begin{equation}
	\label{eq:CP3GOevanRGE}
	- \sum_j 2 K_j^{(1,1)} \frac{\p L_{\widetilde G}^{(1,0)}}{\p K_j^r} = 2 \times \frac{1}{3} g^4 N_c ( n_q + N_c ) L_{\widetilde G} \, .
\end{equation}
Dropping the contribution of finite renormalizations, we obtain the running of the $CP$-odd three-gluon operator in the primed basis in the \msbar{} scheme
\begin{align}
	\label{eq:CP3GORGEmsbar}
	\frac{\p}{\p L_{\widetilde G}'} \left[ \dot L_{\widetilde G}' \right]_2^\text{\msbar} &= \frac{\p}{\p L_{\widetilde G}} \left[ \dot L_{\widetilde G} \right]_2 + 4 \times \frac{1}{18} g^4 N_c ( 3 n_q - 8 N_c ) - 2 \times \frac{1}{3} g^4 N_c ( n_q + N_c )  \nn
		&= 6 e^2 g^2 (n_u \q_u^2 + n_d \q_d^2) + n_q \frac{g^4 \left(N_c^2-9\right)}{3N_c} + \frac{119 g^4 N_c^2}{9} \, ,
\end{align}
which agrees with Ref.~\cite{deVries:2019nsu} for the QCD part. However, this is not the complete RGE of the three-gluon operator: in the \msbar{} scheme, we find a two-loop mixing of the coefficients of evanescent two-quark operators into $L_{\widetilde G}'$, i.e.,
\begin{equation}
	\frac{\p}{\p K_j} \left[ \dot L_{\widetilde G}' \right]_2^\text{\msbar}
\end{equation}
does not vanish for all evanescent-operator coefficients. Since these evanescent operators are generated at one loop from the insertion of $\O_{\widetilde G}'$ into the quark--antiquark--gluon three-point function, the running~\eqref{eq:CP3GORGEmsbar} alone does not correctly resum all next-to-leading logs in the HV scheme. A mixing of the evanescent into the physical sector is not a desirable scheme. Instead, it is common to renormalize the evanescent operators to avoid such a mixing~\cite{Dugan:1990df,Herrlich:1994kh}. This induces an additional contribution to the RGE
\begin{equation}
	\label{eq:CP3GOevanRGEprime}
	- \sum_j 2 K_j^{\prime(1,1)} \frac{\p L_{\widetilde G}^{\prime(1,0)}}{\p K_j^r} = 2 \times \frac{2}{9} g^4 N_c n_q L_{\widetilde G}' \, ,
\end{equation}
which changes the RGE in the primed basis (and in the HV scheme) to Eq.~\eqref{eq:CP3GORGEprime}.

%% file: Paper-dim6.bbl
\providecommand{\href}[2]{#2}\begingroup\raggedright\begin{thebibliography}{100}

\bibitem{Buchmuller:1985jz}
W.~Buchm\"uller and D.~Wyler, Nucl. Phys. B {\bfseries 268}, 621 (1986).

\bibitem{Grzadkowski:2010es}
B.~Grzadkowski, M.~Iskrzynski, M.~Misiak, and J.~Rosiek, JHEP {\bfseries 10},
  085 (2010),
  [\href{https://arxiv.org/abs/1008.4884}{{arXiv:1008.4884~[hep-ph]}}].

\bibitem{Lehman:2014jma}
L.~Lehman, Phys. Rev. D {\bfseries 90}, 125023 (2014),
  [\href{https://arxiv.org/abs/1410.4193}{{arXiv:1410.4193~[hep-ph]}}].

\bibitem{Liao:2016hru}
Y.~Liao and X.-D. Ma, JHEP {\bfseries 11}, 043 (2016),
  [\href{https://arxiv.org/abs/1607.07309}{{arXiv:1607.07309~[hep-ph]}}].

\bibitem{Murphy:2020rsh}
C.~W. Murphy, JHEP {\bfseries 10}, 174 (2020),
  [\href{https://arxiv.org/abs/2005.00059}{{arXiv:2005.00059~[hep-ph]}}].

\bibitem{Li:2020gnx}
H.-L. Li, Z.~Ren, J.~Shu, M.-L. Xiao, J.-H. Yu, and Y.-H. Zheng, Phys. Rev. D
  {\bfseries 104}, 015026 (2021),
  [\href{https://arxiv.org/abs/2005.00008}{{arXiv:2005.00008~[hep-ph]}}].

\bibitem{Liao:2020jmn}
Y.~Liao and X.-D. Ma, JHEP {\bfseries 11}, 152 (2020),
  [\href{https://arxiv.org/abs/2007.08125}{{arXiv:2007.08125~[hep-ph]}}].

\bibitem{Harlander:2023psl}
R.~V. Harlander, T.~Kempkens, and M.~C. Schaaf, Phys. Rev. D {\bfseries 108},
  055020 (2023),
  [\href{https://arxiv.org/abs/2305.06832}{{arXiv:2305.06832~[hep-ph]}}].

\bibitem{Jenkins:2017jig}
E.~E. Jenkins, A.~V. Manohar, and P.~Stoffer, JHEP {\bfseries 03}, 016 (2018),
  [\href{https://arxiv.org/abs/1709.04486}{{arXiv:1709.04486~[hep-ph]}}],
  [Erratum: JHEP {\bf 12}, 043 (2023)].

\bibitem{Liao:2020zyx}
Y.~Liao, X.-D. Ma, and Q.-Y. Wang, JHEP {\bfseries 08}, 162 (2020),
  [\href{https://arxiv.org/abs/2005.08013}{{arXiv:2005.08013~[hep-ph]}}].

\bibitem{Murphy:2020cly}
C.~W. Murphy, JHEP {\bfseries 04}, 101 (2021),
  [\href{https://arxiv.org/abs/2012.13291}{{arXiv:2012.13291~[hep-ph]}}].

\bibitem{Li:2020tsi}
H.-L. Li, Z.~Ren, M.-L. Xiao, J.-H. Yu, and Y.-H. Zheng, JHEP {\bfseries 06},
  138 (2021),
  [\href{https://arxiv.org/abs/2012.09188}{{arXiv:2012.09188~[hep-ph]}}].

\bibitem{Jenkins:2013zja}
E.~E. Jenkins, A.~V. Manohar, and M.~Trott, JHEP {\bfseries 10}, 087 (2013),
  [\href{https://arxiv.org/abs/1308.2627}{{arXiv:1308.2627~[hep-ph]}}].

\bibitem{Jenkins:2013wua}
E.~E. Jenkins, A.~V. Manohar, and M.~Trott, JHEP {\bfseries 01}, 035 (2014),
  [\href{https://arxiv.org/abs/1310.4838}{{arXiv:1310.4838~[hep-ph]}}].

\bibitem{Alonso:2013hga}
R.~Alonso, E.~E. Jenkins, A.~V. Manohar, and M.~Trott, JHEP {\bfseries 04}, 159
  (2014), [\href{https://arxiv.org/abs/1312.2014}{{arXiv:1312.2014~[hep-ph]}}].

\bibitem{Jenkins:2017dyc}
E.~E. Jenkins, A.~V. Manohar, and P.~Stoffer, JHEP {\bfseries 01}, 084 (2018),
  [\href{https://arxiv.org/abs/1711.05270}{{arXiv:1711.05270~[hep-ph]}}],
  [Erratum: JHEP {\bf 12}, 042 (2023)].

\bibitem{Buras:1989xd}
A.~J. Buras and P.~H. Weisz, Nucl. Phys. B {\bfseries 333}, 66 (1990).

\bibitem{Buras:1991jm}
A.~J. Buras, M.~Jamin, M.~E. Lautenbacher, and P.~H. Weisz, Nucl. Phys. B
  {\bfseries 370}, 69 (1992), [Addendum: Nucl.~Phys.~B {\bf 375}, 501 (1992)].

\bibitem{Buras:1992tc}
A.~J. Buras, M.~Jamin, M.~E. Lautenbacher, and P.~H. Weisz, Nucl. Phys. B
  {\bfseries 400}, 37 (1993),
  [\href{https://arxiv.org/abs/hep-ph/9211304}{{arXiv:hep-ph/9211304}}].

\bibitem{Ciuchini:1993vr}
M.~Ciuchini, E.~Franco, G.~Martinelli, and L.~Reina, Nucl. Phys. B {\bfseries
  415}, 403 (1994),
  [\href{https://arxiv.org/abs/hep-ph/9304257}{{arXiv:hep-ph/9304257}}].

\bibitem{Ciuchini:1993fk}
M.~Ciuchini, E.~Franco, L.~Reina, and L.~Silvestrini, Nucl. Phys. B {\bfseries
  421}, 41 (1994),
  [\href{https://arxiv.org/abs/hep-ph/9311357}{{arXiv:hep-ph/9311357}}].

\bibitem{Buchalla:1995vs}
G.~Buchalla, A.~J. Buras, and M.~E. Lautenbacher, Rev. Mod. Phys. {\bfseries
  68}, 1125 (1996),
  [\href{https://arxiv.org/abs/hep-ph/9512380}{{arXiv:hep-ph/9512380}}].

\bibitem{Chetyrkin:1997gb}
K.~G. Chetyrkin, M.~Misiak, and M.~M\"unz, Nucl. Phys. B {\bfseries 520}, 279
  (1998),
  [\href{https://arxiv.org/abs/hep-ph/9711280}{{arXiv:hep-ph/9711280}}].

\bibitem{Buras:2000if}
A.~J. Buras, M.~Misiak, and J.~Urban, Nucl. Phys. B {\bfseries 586}, 397
  (2000),
  [\href{https://arxiv.org/abs/hep-ph/0005183}{{arXiv:hep-ph/0005183}}].

\bibitem{Bobeth:2003at}
C.~Bobeth, P.~Gambino, M.~Gorbahn, and U.~Haisch, JHEP {\bfseries 04}, 071
  (2004),
  [\href{https://arxiv.org/abs/hep-ph/0312090}{{arXiv:hep-ph/0312090}}].

\bibitem{Gorbahn:2004my}
M.~Gorbahn and U.~Haisch, Nucl. Phys. B {\bfseries 713}, 291 (2005),
  [\href{https://arxiv.org/abs/hep-ph/0411071}{{arXiv:hep-ph/0411071}}].

\bibitem{Huber:2005ig}
T.~Huber, E.~Lunghi, M.~Misiak, and D.~Wyler, Nucl. Phys. B {\bfseries 740},
  105 (2006),
  [\href{https://arxiv.org/abs/hep-ph/0512066}{{arXiv:hep-ph/0512066}}].

\bibitem{Gorbahn:2005sa}
M.~Gorbahn, U.~Haisch, and M.~Misiak, Phys. Rev. Lett. {\bfseries 95}, 102004
  (2005),
  [\href{https://arxiv.org/abs/hep-ph/0504194}{{arXiv:hep-ph/0504194}}].

\bibitem{Czakon:2006ss}
M.~Czakon, U.~Haisch, and M.~Misiak, JHEP {\bfseries 03}, 008 (2007),
  [\href{https://arxiv.org/abs/hep-ph/0612329}{{arXiv:hep-ph/0612329}}].

\bibitem{Aebischer:2017gaw}
J.~Aebischer, M.~Fael, C.~Greub, and J.~Virto, JHEP {\bfseries 09}, 158 (2017),
  [\href{https://arxiv.org/abs/1704.06639}{{arXiv:1704.06639~[hep-ph]}}].

\bibitem{Panico:2018hal}
G.~Panico, A.~Pomarol, and M.~Riembau, JHEP {\bfseries 04}, 090 (2019),
  [\href{https://arxiv.org/abs/1810.09413}{{arXiv:1810.09413~[hep-ph]}}].

\bibitem{Morell:2024aml}
P.~Morell and J.~Virto, JHEP {\bfseries 04}, 105 (2024),
  [\href{https://arxiv.org/abs/2402.00249}{{arXiv:2402.00249~[hep-ph]}}].

\bibitem{Carmona:2021xtq}
A.~Carmona, A.~Lazopoulos, P.~Olgoso, and J.~Santiago, SciPost Phys. {\bfseries
  12}, 198 (2022),
  [\href{https://arxiv.org/abs/2112.10787}{{arXiv:2112.10787~[hep-ph]}}].

\bibitem{Fuentes-Martin:2022jrf}
J.~Fuentes-Mart\'\i{}n, M.~K\"onig, J.~Pag\`es, A.~E. Thomsen, and F.~Wilsch,
  Eur. Phys. J. C {\bfseries 83}, 662 (2023),
  [\href{https://arxiv.org/abs/2212.04510}{{arXiv:2212.04510~[hep-ph]}}].

\bibitem{Fuentes-Martin:2023ljp}
J.~Fuentes-Mart\'\i{}n, A.~Palavri\'c, and A.~E. Thomsen, Phys. Lett. B
  {\bfseries 851}, 138557 (2024),
  [\href{https://arxiv.org/abs/2311.13630}{{arXiv:2311.13630~[hep-ph]}}].

\bibitem{Aebischer:2023nnv}
L.~Allwicher {\em et~al.}, Eur. Phys. J. C {\bfseries 84}, 170 (2024),
  [\href{https://arxiv.org/abs/2307.08745}{{arXiv:2307.08745~[hep-ph]}}].

\bibitem{Thomsen:2024abg}
A.~E. Thomsen, JHEP {\bfseries 12}, 185 (2024),
  [\href{https://arxiv.org/abs/2404.11640}{{arXiv:2404.11640~[hep-ph]}}].

\bibitem{Dekens:2019ept}
W.~Dekens and P.~Stoffer, JHEP {\bfseries 10}, 197 (2019),
  [\href{https://arxiv.org/abs/1908.05295}{{arXiv:1908.05295~[hep-ph]}}],
  [Erratum: JHEP {\bf 11}, 148 (2022)].

\bibitem{Gorbahn:2016uoy}
M.~Gorbahn and U.~Haisch, JHEP {\bfseries 10}, 094 (2016),
  [\href{https://arxiv.org/abs/1607.03773}{{arXiv:1607.03773~[hep-ph]}}].

\bibitem{deVries:2019nsu}
J.~de~Vries, G.~Falcioni, F.~Herzog, and B.~Ruijl, Phys. Rev. D {\bfseries
  102}, 016010 (2020),
  [\href{https://arxiv.org/abs/1907.04923}{{arXiv:1907.04923~[hep-ph]}}].

\bibitem{Bern:2020ikv}
Z.~Bern, J.~Parra-Martinez, and E.~Sawyer, JHEP {\bfseries 10}, 211 (2020),
  [\href{https://arxiv.org/abs/2005.12917}{{arXiv:2005.12917~[hep-ph]}}].

\bibitem{Aebischer:2022anv}
J.~Aebischer, A.~J. Buras, and J.~Kumar, Phys. Rev. D {\bfseries 106}, 035003
  (2022),
  [\href{https://arxiv.org/abs/2203.11224}{{arXiv:2203.11224~[hep-ph]}}].

\bibitem{Fuentes-Martin:2022vvu}
J.~Fuentes-Mart\'\i{}n, M.~K\"onig, J.~Pag\`es, A.~E. Thomsen, and F.~Wilsch,
  JHEP {\bfseries 02}, 031 (2023),
  [\href{https://arxiv.org/abs/2211.09144}{{arXiv:2211.09144~[hep-ph]}}].

\bibitem{Aebischer:2023djt}
J.~Aebischer, M.~Pesut, and Z.~Polonsky, JHEP {\bfseries 01}, 060 (2024),
  [\href{https://arxiv.org/abs/2306.16449}{{arXiv:2306.16449~[hep-ph]}}].

\bibitem{Jenkins:2023rtg}
E.~E. Jenkins, A.~V. Manohar, L.~Naterop, and J.~Pag\`es, JHEP {\bfseries 12},
  165 (2023),
  [\href{https://arxiv.org/abs/2308.06315}{{arXiv:2308.06315~[hep-ph]}}].

\bibitem{Jenkins:2023bls}
E.~E. Jenkins, A.~V. Manohar, L.~Naterop, and J.~Pag\`es, JHEP {\bfseries 02},
  131 (2024),
  [\href{https://arxiv.org/abs/2310.19883}{{arXiv:2310.19883~[hep-ph]}}].

\bibitem{Naterop:2023dek}
L.~Naterop and P.~Stoffer, JHEP {\bfseries 02}, 068 (2024),
  [\href{https://arxiv.org/abs/2310.13051}{{arXiv:2310.13051~[hep-ph]}}].

\bibitem{DiNoi:2023ygk}
S.~Di~Noi, R.~Gr\"ober, G.~Heinrich, J.~Lang, and M.~Vitti, Phys. Rev. D
  {\bfseries 109}, 095024 (2024),
  [\href{https://arxiv.org/abs/2310.18221}{{arXiv:2310.18221~[hep-ph]}}].

\bibitem{Aebischer:2024xnf}
J.~Aebischer, M.~Pesut, and Z.~Polonsky, Eur. Phys. J. C {\bfseries 84}, 750
  (2024),
  [\href{https://arxiv.org/abs/2401.16904}{{arXiv:2401.16904~[hep-ph]}}].

\bibitem{Manohar:2024xbh}
A.~V. Manohar, J.~Pag\`es, and J.~Roosmale~Nepveu, JHEP {\bfseries 05}, 018
  (2024),
  [\href{https://arxiv.org/abs/2402.08715}{{arXiv:2402.08715~[hep-ph]}}].

\bibitem{DiNoi:2024ajj}
S.~Di~Noi, R.~Gr\"ober, and M.~K. Mandal, JHEP {\bfseries 12}, 220 (2025),
  [\href{https://arxiv.org/abs/2408.03252}{{arXiv:2408.03252~[hep-ph]}}].

\bibitem{Born:2024mgz}
L.~Born, J.~Fuentes-Mart{\'\i}n, S.~Kvedarait{\.{e}}, and A.~E. Thomsen, JHEP
  {\bfseries 05}, 121 (2025),
  [\href{https://arxiv.org/abs/2410.07320}{{arXiv:2410.07320~[hep-ph]}}].

\bibitem{Naterop:2024cfx}
L.~Naterop and P.~Stoffer, JHEP {\bfseries 06}, 007 (2025),
  [\href{https://arxiv.org/abs/2412.13251}{{arXiv:2412.13251~[hep-ph]}}].

\bibitem{Fuentes-Martin:2024agf}
J.~Fuentes-Mart{\'\i}n, A.~Moreno-S{\'a}nchez, A.~Palavri{\'c}, and A.~E.
  Thomsen, JHEP {\bfseries 08}, 099 (2025),
  [\href{https://arxiv.org/abs/2412.12270}{{arXiv:2412.12270~[hep-ph]}}].

\bibitem{Aebischer:2025hsx}
J.~Aebischer, P.~Morell, M.~Pesut, and J.~Virto,
  \href{https://arxiv.org/abs/2501.08384}{{arXiv:2501.08384~[hep-ph]}}.

\bibitem{Duhr:2025zqw}
C.~Duhr, A.~Vasquez, G.~Ventura, and E.~Vryonidou, JHEP {\bfseries 07}, 160
  (2025),
  [\href{https://arxiv.org/abs/2503.01954}{{arXiv:2503.01954~[hep-ph]}}].

\bibitem{Haisch:2025lvd}
U.~Haisch, JHEP {\bfseries 06}, 004 (2025),
  [\href{https://arxiv.org/abs/2503.06249}{{arXiv:2503.06249~[hep-ph]}}].

\bibitem{Naterop:2025lzc}
L.~Naterop and P.~Stoffer, JHEP {\bfseries 07}, 237 (2025),
  [\href{https://arxiv.org/abs/2505.03871}{{arXiv:2505.03871~[hep-ph]}}].

\bibitem{Haisch:2025vqj}
U.~Haisch and M.~Niggetiedt,
  \href{https://arxiv.org/abs/2507.20803}{{arXiv:2507.20803~[hep-ph]}}.

\bibitem{Duhr:2025yor}
C.~Duhr, G.~Ventura, and E.~Vryonidou, JHEP {\bfseries 11}, 046 (2025),
  [\href{https://arxiv.org/abs/2508.04500}{{arXiv:2508.04500~[hep-ph]}}].

\bibitem{Banik:2025wpi}
S.~Banik, A.~Crivellin, L.~Naterop, and P.~Stoffer, JHEP {\bfseries 02}, 017
  (2026),
  [\href{https://arxiv.org/abs/2510.08682}{{arXiv:2510.08682~[hep-ph]}}].

\bibitem{Alarcon:2022ero}
R.~Alarcon {\em et~al.}, ``{Electric dipole moments and the search for new
  physics},'' in {\em {Snowmass 2021}}.
\newblock 3, 2022.
\newblock \href{https://arxiv.org/abs/2203.08103}{{arXiv:2203.08103~[hep-ph]}}.

\bibitem{Jegerlehner:2000dz}
F.~Jegerlehner, Eur. Phys. J. C {\bfseries 18}, 673 (2001),
  [\href{https://arxiv.org/abs/hep-th/0005255}{{arXiv:hep-th/0005255}}].

\bibitem{tHooft:1972tcz}
G.~'t~Hooft and M.~J.~G. Veltman, Nucl. Phys. B {\bfseries 44}, 189 (1972).

\bibitem{Breitenlohner:1975hg}
P.~Breitenlohner and D.~Maison, Commun. Math. Phys. {\bfseries 52}, 39 (1977).

\bibitem{Breitenlohner:1976te}
P.~Breitenlohner and D.~Maison, Commun. Math. Phys. {\bfseries 52}, 55 (1977).

\bibitem{Breitenlohner:1977hr}
P.~Breitenlohner and D.~Maison, Commun. Math. Phys. {\bfseries 52}, 11 (1977).

\bibitem{Schubert:1988ke}
C.~Schubert, Nucl. Phys. B {\bfseries 323}, 478 (1989).

\bibitem{Ferrari:1994ct}
R.~Ferrari, A.~Le~Yaouanc, L.~Oliver, and J.~C. Raynal, Phys. Rev. D {\bfseries
  52}, 3036 (1995).

\bibitem{Cornella:2022hkc}
C.~Cornella, F.~Feruglio, and L.~Vecchi, JHEP {\bfseries 02}, 244 (2023),
  [\href{https://arxiv.org/abs/2205.10381}{{arXiv:2205.10381~[hep-ph]}}].

\bibitem{Belusca-Maito:2020ala}
H.~B\'elusca-Ma\"\i{}to, A.~Ilakovac, M.~Ma\dj{}or-Bo\v{z}inovi\'c, and
  D.~St\"ockinger, JHEP {\bfseries 08}, 024 (2020),
  [\href{https://arxiv.org/abs/2004.14398}{{arXiv:2004.14398~[hep-ph]}}].

\bibitem{Belusca-Maito:2021lnk}
H.~B\'elusca-Ma\"\i{}to, A.~Ilakovac, P.~K\"uhler,
  M.~Ma\dj{}or-Bo\v{z}inovi\'c, and D.~St\"ockinger, JHEP {\bfseries 11}, 159
  (2021),
  [\href{https://arxiv.org/abs/2109.11042}{{arXiv:2109.11042~[hep-ph]}}].

\bibitem{Belusca-Maito:2023wah}
H.~B\'elusca-Ma\"\i{}to, A.~Ilakovac, P.~K\"uhler,
  M.~Ma\dj{}or-Bo\v{z}inovi\'c, D.~St\"ockinger, and M.~Wei\ss{}wange, Symmetry
  {\bfseries 15}, 622 (2023),
  [\href{https://arxiv.org/abs/2303.09120}{{arXiv:2303.09120~[hep-ph]}}].

\bibitem{Stockinger:2023ndm}
D.~St\"ockinger and M.~Wei\ss{}wange, JHEP {\bfseries 02}, 139 (2024),
  [\href{https://arxiv.org/abs/2312.11291}{{arXiv:2312.11291~[hep-ph]}}].

\bibitem{OlgosoRuiz:2024dzq}
P.~Olgoso~Ruiz and L.~Vecchi, JHEP {\bfseries 12}, 080 (2024),
  [\href{https://arxiv.org/abs/2406.17013}{{arXiv:2406.17013~[hep-ph]}}].

\bibitem{Ebert:2024xpy}
P.~L. Ebert, P.~K{\"u}hler, D.~St{\"o}ckinger, and M.~Wei{\ss}wange, JHEP
  {\bfseries 01}, 114 (2025),
  [\href{https://arxiv.org/abs/2411.02543}{{arXiv:2411.02543~[hep-ph]}}].

\bibitem{DiNoi:2025uan}
S.~Di~Noi, R.~Gr{\"o}ber, and P.~Olgoso, JHEP {\bfseries 09}, 027 (2025),
  [\href{https://arxiv.org/abs/2504.00112}{{arXiv:2504.00112~[hep-ph]}}].

\bibitem{vonManteuffel:2025swv}
A.~von Manteuffel, D.~St{\"o}ckinger, and M.~Wei{\ss}wange, JHEP {\bfseries
  08}, 088 (2025),
  [\href{https://arxiv.org/abs/2506.12253}{{arXiv:2506.12253~[hep-ph]}}].

\bibitem{Fuentes-Martin:2025meq}
J.~Fuentes-Mart{\'\i}n, A.~Moreno-S{\'a}nchez, and A.~E. Thomsen,
  \href{https://arxiv.org/abs/2507.19589}{{arXiv:2507.19589~[hep-ph]}}.

\bibitem{Kreimer:1989ke}
D.~Kreimer, Phys. Lett. B {\bfseries 237}, 59 (1990).

\bibitem{Korner:1991sx}
J.~G. Korner, D.~Kreimer, and K.~Schilcher, Z. Phys. C {\bfseries 54}, 503
  (1992).

\bibitem{Chen:2023lus}
L.~Chen, JHEP {\bfseries 11}, 030 (2023),
  [\href{https://arxiv.org/abs/2304.13814}{{arXiv:2304.13814~[hep-ph]}}].

\bibitem{Chen:2024hlv}
L.~Chen, JHEP {\bfseries 05}, 109 (2025),
  [\href{https://arxiv.org/abs/2409.08099}{{arXiv:2409.08099~[hep-ph]}}].

\bibitem{Abbott:1980hw}
L.~F. Abbott, Nucl. Phys. B {\bfseries 185}, 189 (1981).

\bibitem{Abbott:1983zw}
L.~F. Abbott, M.~T. Grisaru, and R.~K. Schaefer, Nucl. Phys. B {\bfseries 229},
  372 (1983).

\bibitem{Chetyrkin:1997fm}
K.~G. Chetyrkin, M.~Misiak, and M.~M\"unz, Nucl. Phys. B {\bfseries 518}, 473
  (1998),
  [\href{https://arxiv.org/abs/hep-ph/9711266}{{arXiv:hep-ph/9711266}}].

\bibitem{tHooft:1973mfk}
G.~'t~Hooft, Nucl. Phys. B {\bfseries 61}, 455 (1973).

\bibitem{Bednyakov:2014pia}
A.~V. Bednyakov, A.~F. Pikelner, and V.~N. Velizhanin, Phys. Lett. B {\bfseries
  737}, 129 (2014),
  [\href{https://arxiv.org/abs/1406.7171}{{arXiv:1406.7171~[hep-ph]}}].

\bibitem{Herren:2017uxn}
F.~Herren, L.~Mihaila, and M.~Steinhauser, Phys. Rev. D {\bfseries 97}, 015016
  (2018),
  [\href{https://arxiv.org/abs/1712.06614}{{arXiv:1712.06614~[hep-ph]}}],
  [Erratum: Phys.~Rev.~D {\bf 101}, 079903 (2020)].

\bibitem{Herren:2021yur}
F.~Herren and A.~E. Thomsen, JHEP {\bfseries 06}, 116 (2021),
  [\href{https://arxiv.org/abs/2104.07037}{{arXiv:2104.07037~[hep-th]}}].

\bibitem{Zhang:2025ywe}
D.~Zhang, JHEP {\bfseries 06}, 106 (2025),
  [\href{https://arxiv.org/abs/2504.00792}{{arXiv:2504.00792~[hep-ph]}}].

\bibitem{Jenkins:2009dy}
E.~E. Jenkins and A.~V. Manohar, JHEP {\bfseries 10}, 094 (2009),
  [\href{https://arxiv.org/abs/0907.4763}{{arXiv:0907.4763~[hep-ph]}}].

\bibitem{Nogueira:1991ex}
P.~Nogueira, J. Comput. Phys. {\bfseries 105}, 279 (1993).

\bibitem{Vermaseren:2000nd}
J.~A.~M. Vermaseren,
  \href{https://arxiv.org/abs/math-ph/0010025}{{arXiv:math-ph/0010025}}.

\bibitem{Ruijl:2017dtg}
B.~Ruijl, T.~Ueda, and J.~Vermaseren,
  \href{https://arxiv.org/abs/1707.06453}{{arXiv:1707.06453~[hep-ph]}}.

\bibitem{Altarelli:1980fi}
G.~Altarelli, G.~Curci, G.~Martinelli, and S.~Petrarca, Nucl. Phys. B
  {\bfseries 187}, 461 (1981).

\bibitem{Collins:2005nj}
J.~C. Collins, A.~V. Manohar, and M.~B. Wise, Phys. Rev. D {\bfseries 73},
  105019 (2006),
  [\href{https://arxiv.org/abs/hep-th/0512187}{{arXiv:hep-th/0512187}}].

\bibitem{Naterop:2025factorizable}
L.~Naterop, in preparation.

\bibitem{Khatsimovsky:1987fr}
V.~M. Khatsimovsky, I.~B. Khriplovich, and A.~S. Yelkhovsky, Annals Phys.
  {\bfseries 186}, 1 (1988).

\bibitem{Buhler:2023gsg}
J.~B\"uhler and P.~Stoffer, JHEP {\bfseries 08}, 194 (2023),
  [\href{https://arxiv.org/abs/2304.00985}{{arXiv:2304.00985~[hep-lat]}}].

\bibitem{Larin:1993tq}
S.~A. Larin, Phys. Lett. B {\bfseries 303}, 113 (1993),
  [\href{https://arxiv.org/abs/hep-ph/9302240}{{arXiv:hep-ph/9302240}}].

\bibitem{Dugan:1990df}
M.~J. Dugan and B.~Grinstein, Phys. Lett. B {\bfseries 256}, 239 (1991).

\bibitem{Herrlich:1994kh}
S.~Herrlich and U.~Nierste, Nucl. Phys. B {\bfseries 455}, 39 (1995),
  [\href{https://arxiv.org/abs/hep-ph/9412375}{{arXiv:hep-ph/9412375}}].

\bibitem{Cirigliano:2020msr}
V.~Cirigliano, E.~Mereghetti, and P.~Stoffer, JHEP {\bfseries 09}, 094 (2020),
  [\href{https://arxiv.org/abs/2004.03576}{{arXiv:2004.03576~[hep-ph]}}].

\bibitem{Crosas:2023anw}
{\`O}.~L. Crosas, C.~J. Monahan, M.~D. Rizik, A.~Shindler, and P.~Stoffer,
  Phys. Lett. B {\bfseries 847}, 138301 (2023),
  [\href{https://arxiv.org/abs/2308.16221}{{arXiv:2308.16221~[hep-lat]}}].

\bibitem{Naterop:2025inPrep}
L.~Naterop and P.~Stoffer, in preparation.

\bibitem{DiNoi:2025arz}
S.~Di~Noi and R.~Gr{\"o}ber, Phys. Lett. B {\bfseries 869}, 139878 (2025),
  [\href{https://arxiv.org/abs/2507.10295}{{arXiv:2507.10295~[hep-ph]}}].

\end{thebibliography}\endgroup
